\documentclass[fleqn,10pt]{wlscirep}

\usepackage[sort&compress,numbers]{natbib}
\setcitestyle{numbers,square}

\usepackage[absolute,overlay]{textpos} 

\usepackage[utf8]{inputenc}
\usepackage[T1]{fontenc}
\usepackage{bm}
\usepackage[textsize=tiny,backgroundcolor=yellow]{todonotes}
\usepackage[nolist]{acronym}
\usepackage{xcolor}
\usepackage[noabbrev]{cleveref}
\usepackage{xspace}
\usepackage{braket}
\usepackage{colortbl}
\usepackage{blkarray}
\usepackage[normalem]{ulem}
\usepackage{lineno}

\usepackage{float}
\usepackage{amsmath}      
\usepackage{amssymb} 
\usepackage{booktabs}  
\usepackage{xcolor} 
\usepackage{graphicx} 
\usepackage{caption}  
\usepackage{subcaption}
\usepackage{dsfont}

\usepackage{placeins}

\definecolor{amethyst}{rgb}{0.6, 0.4, 0.8}

\newcommand{\SU}[1]{\ensuremath{\mathrm{SU}(#1)}}
\newcommand{\U}[1]{\ensuremath{\mathrm{U}(#1)}}
\newcommand{\CP}{\ensuremath{CP}\xspace}
\newcommand{\ee}{\mathrm{e}}
\newcommand{\ii}{\hskip0.1ex\mathrm{i}\hskip0.1ex}
\newcommand{\rep}[1]{\ensuremath{\boldsymbol{#1}}}

\newcommand{\Id}{\ensuremath{\mathds{1}}}
\makeatletter
\newcommand\footnoteref[1]{\protected@xdef\@thefnmark{\ref{#1}}\@footnotemark}
\makeatother



\usepackage{nameref} 

\newcommand{\myref}[1]{``\nameref*{#1}''}

\crefname{table}{Tab.}{Tabs.}
\Crefname{table}{Tab.}{Tabs.}

\crefname{figure}{Fig.}{Figs.}
\Crefname{figure}{Fig.}{Figs.}

\title{Towards AI-assisted Neutrino Flavor Theory Design}

\author[1,$\alpha$]{Jason Benjamin Baretz}
\author[1,2,$\beta$]{Max Fieg}
\author[3, $\gamma$]{Vijay Ganesh}
\author[1,3,4,$\delta$,*]{Aishik Ghosh}
\author[1, $\epsilon$]{V. Knapp-P\'{e}rez}
\author[1, $\zeta$,*]{Jake Rudolph}
\author[1, $\eta$]{Daniel Whiteson}

\affil[1]{Department of Physics and Astronomy, University\ of\
  California, Irvine, CA\ 92697}
  \affil[2]{Particle Theory Department, Fermilab, Batavia, IL 60510}

\affil[3]{Georgia Institute of Technology, Atlanta, GA 30332}
\affil[4]{Physics Division, Lawrence Berkeley National Laboratory, Berkeley, CA 94720}

\affil[$\alpha$]{e-mail: jbaretz@uci.edu}
\affil[$\beta$]{e-mail: mfieg@uci.edu}
\affil[$\gamma$]{e-mail: vganesh@gatech.edu}
\affil[$\delta$]{e-mail: aishikghosh@physics.gatech.edu}
\affil[$\epsilon$]{e-mail: vknapppe@uci.edu}
\affil[$\zeta$]{e-mail: jirudolp@uci.edu}
\affil[$\eta$]{e-mail: daniel@uci.edu}
\affil[*]{Corresponding authors}
\affil[ ]{}
\affil[ ]{Authors are listed alphabetically.}

\begin{abstract}

Particle physics theories, such as those which explain neutrino flavor mixing, arise from a vast landscape of model-building possibilities. A model's construction typically relies on the intuition of theorists. It also requires considerable effort to identify appropriate symmetry groups, assign field representations, and extract predictions for comparison with experimental data. We develop \ac{AMBer}, a framework in which a reinforcement learning agent interacts with a streamlined physics software pipeline to search these spaces efficiently. \ac{AMBer} selects symmetry groups, particle content, and group representation assignments to construct models while minimizing the number of free parameters introduced. We validate our approach in well-studied regions of theory space and extend the exploration to a previously unexamined symmetry group. While demonstrated in the context of neutrino flavor theories, this approach of reinforcement learning with physics software feedback may be extended to other theoretical model-building problems in the future.

\end{abstract}

\begin{document}

\setlength{\TPHorizModule}{1cm}
\setlength{\TPVertModule}{1cm}

\begin{textblock*}{5cm}(15cm,1.5cm)
  \raggedleft \text{UCI-TR-2025-05}
\end{textblock*}

\flushbottom
\maketitle

\thispagestyle{empty}

\section{Introduction}
\label{sec:Introduction}

Particle physics seeks answers to basic questions about the nature of matter and its interactions. Answers are not in the form of raw experimental results, but are expressed as a {\it model}~\cite{fox2022tf08snowmassreportbsm} of particles and their interactions that match observations and make predictions for previously unexplained phenomena. In the case of neutrino oscillations, for example, a successful model should explain the observed neutrino masses and oscillations and minimize additional theoretical elements to maximize predictive power~\cite{feruglio2021leptonflavoursymmetries,Almumin:2022rml,degouvêa2022theoryneutrinophysics}.

Building such a model is far from trivial, as the space of theoretical options is vast and high-dimensional, but constrained by experimental results. At the same time, the twin goals of achieving agreement with experiment and minimizing degrees of freedom are often in tension. Complicating matters further, theory spaces themselves are not conducive to systematic searches, as small changes in model content or structure can yield dramatic changes in quality. Finally, there is the computational challenge that an individual model takes significant work to evaluate: proposing a set of symmetries and fields, assigning representations of these symmetries to the fields, and performing intricate calculations to determine the predictions of a model.

Typical approaches rely on theorists to use their experience and inspiration to select and evaluate tailored theories. In the context of neutrino models, the result is a large space that is expensive to explore, containing vast regions which have not been thoroughly considered. Within these spaces may lie valuable, unexplored theories.

Neutrinos in the \ac{SM} are purely left-handed massless states. Thus, due to the constraints of renormalizability and gauge invariance, they cannot acquire mass through the Higgs mechanism within the \ac{SM}. However, the discovery of neutrino oscillations~\cite{Super-Kamiokande:1998kpq,SNO:2002tuh,Cleveland:1998nv,K2K:2002icj} demonstrates that neutrinos must have small, nonzero masses, in direct conflict with the \ac{SM}. This discovery points to physics beyond the SM to explain neutrino masses, which may also explain the flavor structure observed in the SM. Neutrino flavor models seek to explain neutrino masses and the structure of lepton mixing through beyondd the \ac{SM} mechanisms, symmetries, and particles. 

\begin{figure}[t!]
    \centering
    \includegraphics[width=0.9\textwidth]{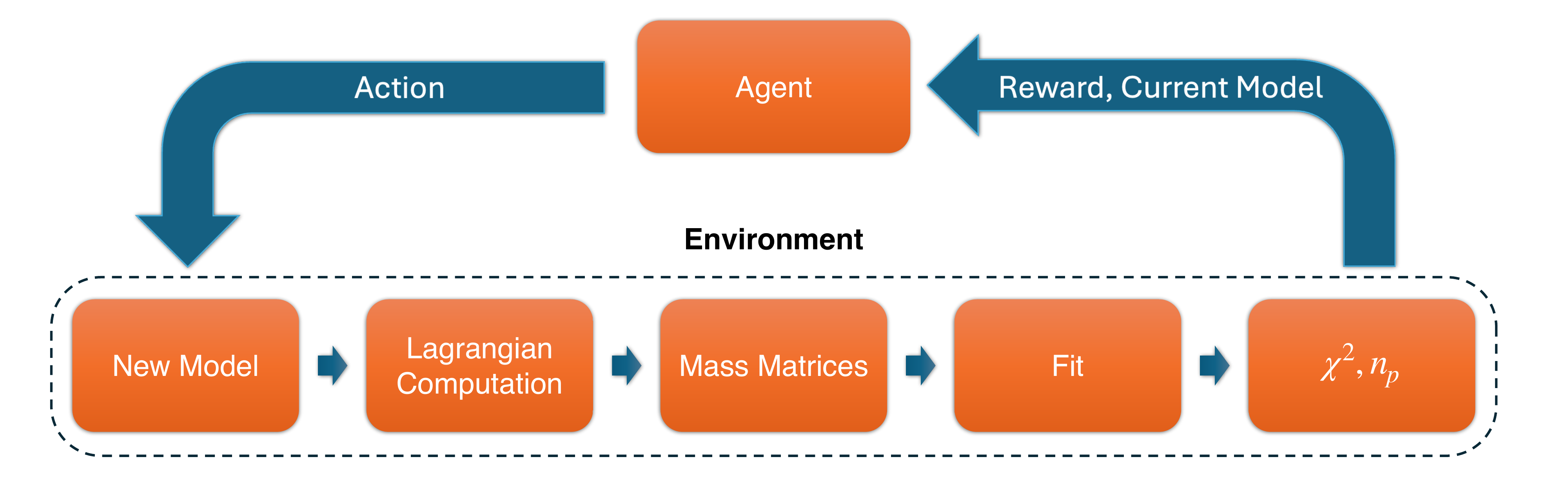}
    \caption{\textbf{Diagram illustrating the reinforcement learning agent, \ac{AMBer}}. \ac{AMBer} searches the space of models, taking actions to modify the model. Each model is then evaluated using a pipeline of physics software, which produces a reward depending on the $\chi^2$ of the fit to data and the number of model parameters. The reward and model inform the agent's selection of the next action. This structure could be generalized to any model-building task by defining the space of models, providing the agent with the physics software necessary to calculate predictions, and designing a reward parameterizing the user's preferences for output models. The result of this learning process is a set of models matching those preferences, which a physicist can use as a basis for further study.
    }
    \label{fig:ConceptDiagram}
\end{figure}

This paper introduces a general-purpose strategy to efficiently search the space of neutrino flavor theories. An optimized scientific software pipeline is developed and interfaced with machine learning to efficiently explore the space and evaluate individual models, reminiscent of automated theorem proving work in mathematics~\cite{10.1007/978-3-540-71070-7_37,jakubuv_et_al:LIPIcs.ITP.2019.34,Blaauwbroek2024}. For the context of neutrino flavor model-building, the scientific software constructs the Lagrangian, extracts the mass matrices and fits model parameters to data. We link this pipeline to a \ac{RL} strategy to intelligently search the space, hunting for models which satisfy the conditions of minimal additional parameters and accurate description of observations. There have been previous applications of RL in particle physics~\cite{Halverson_2019,Nishimura:2020nre, Harvey:2021oue, Nishimura:2024apb, Wojcik:2024lfy, Carta:2025asr}, which focus on minimizing the RL agent's interaction with scientific software. In contrast, we embrace the need to integrate scientific software into the learning loop, giving the agent the same tools as a theorist so it can focus on what it does best: finding complex relationships in high-dimensional spaces. Our approach is summarized in \Cref{fig:ConceptDiagram}. 

While our approach may scale effectively to many domains, we demonstrate it in the context of finding \ac{MSSM}-like seesaw  models to describe observed neutrino oscillations and masses using non-Abelian flavor symmetries broken via a scalar acquiring a \ac{VEV}~\cite{Altarelli_2010}. To verify our framework, it is first applied to the well-explored space of the non-Abelian discrete group $A_4$~\cite{Ma:2001dn,Altarelli:2005yx}, where it rediscovers model patterns previously identified in the literature. We also unleash it on the larger, unexplored space of the group $T_{19} = \mathbb{Z}_{19} \rtimes \mathbb{Z}_{3}$, identifying multiple promising models. Both searches illustrate \ac{AMBer}'s ability to find compelling models worthy of further study. 

Within the models found by AMBer, there could be a dark matter candidate, a mechanism for leptogenesis, or a component of a UV complete theory. AMBer's value is its ability to filter through an unexplored space of flavor symmetries to provide a first pass of patterns and particle representations that reproduce observables. This eliminates the time consuming process of manually tuning these particle representations and fitting parameters to current measurements. We will therefore refer to the models found by AMBer that fit observations with sufficiently few free parameters as “filtered models”.

\section{Methods}
\label{Sec:Methods}

\subsection{Neutrino Flavor Model-Building}
\label{Subsec:NeutrinoModelBuilding}
In this subsection, we outline a common method for neutrino flavor model-building, beginning with the choice of a mass mechanism. A flavor symmetry group is selected to increase predictivity. Next, the particle content is defined and the irreducible representation assignment is chosen. The Lagrangian is constructed following the constraints imposed by the symmetry group. Scalar fields, the  so-called flavons, may be included to break the flavor symmetry through \acp{VEV}. It is assumed that the \acp{VEV} are acquired from an appropriate scalar potential which we do not specify. Finally, the mass matrices are extracted and a fit of the free parameters to experimental data is performed.

A common mechanism to explain neutrino mass is the type-1 seesaw mechanism~\cite{Minkowski:1977sc,Yanagida:1979as,Glashow:1979nm,Gell-Mann:1979vob}, which hypothesizes at least two right-handed neutrinos $N_i$ which have no \ac{SM} gauge interactions. The Lagrangian is given by
\begin{equation}\label{eq:LSee-saw}
    -\mathcal{L}_{\mathrm{see-saw}} ~= ~g^{ij}\tilde{H}\bar{L}^{i}N^{j} + M_M^{ij}\bar{N}^{i,c}N^{j}~+~\mathrm{h.c}\;,
\end{equation}
where $g^{ij}$ is a $3\times 3$ dimensionless coupling matrix, $H$ is the \ac{SM} Higgs, $L^{i}$ are the $\SU{2}_L$ \ac{SM} lepton doublets and $M_M^{ij}$ is the Majorana mass matrix for the heavy right-handed neutrinos $N^i$. Once the Higgs acquires a \ac{VEV} $\braket{h} = v$, the neutrinos  acquire a Dirac mass $M_D^{ij} = v g^{ij}$. Integrating out the heavy degrees of freedom $N^i$ gives the \ac{SM} neutrinos an effective mass of
\begin{equation}\label{eq:Mnu}
    M_\nu  ~=~ - M_D^{T} M_M^{-1} M_D\;.
\end{equation}
The three mass matrices $M_C$, $M_D$ and $M_M$ are $3\times3$ complex matrices resulting in a total of 54 real parameters. Their values are not predicted but must be determined by comparison to experimental observables.  However, there are only 12 physical observables: three charged lepton masses, three neutrino masses, three neutrino mixing angles, one \CP violating phase, and two Majorana phases, under the assumption that neutrinos are Majorana. Of these parameters, only nine have been experimentally measured and seven remain after factoring out an overall mass scale; see the "$A_4$ Properties" subsubsection in the "Supplementary Methods".
Models with more independent parameters than observables are not sufficiently constrained and are thus more descriptive rather than predictive.
The number of parameters can be reduced by applying a symmetry requirement, such as a discrete non-Abelian flavor symmetry~\cite{Ishimori:2010au,Feruglio:2019ybq}, which imposes constraints to reduce the number of independent parameters. However, such a symmetry must be broken at some scale as there are 3 observed generations of leptons with distinct masses and mixings. This breaking is typically mediated by flavon fields, scalars that are charged under the flavor symmetry and acquire a non-zero \ac{VEV}. In this work, it is assumed that the \ac{VEV}s are acquired from an appropriate scalar potential which will remain unspecified.

One of the most well-studied flavor symmetry groups is the alternating group of order four, $A_4$~\cite{Ma:2001dn,Altarelli:2005yx,Ding:2011gt}; see "$A_4$ Properties" subsubsection in the "Supplementary Methods". It is the smallest finite group with a triplet irreducible representation. As a result, the three generations of leptons can be naturally accommodated within a single multiplet.
Here, $A_4$-type models are considered, together with models based on the group $T_{19}$; see Supplementary Tab. 2. The latter is the smallest finite subgroup of $\U{3}$~\cite{Ludl:2010bj} which has not yet been explored in the context of flavor model-building. The symmetry is usually cast within an \ac{MSSM}-like model,  which includes two Higgs $\SU{2}_L$ doublets $H_u$ and $H_d$. There are also three right-handed neutrinos, and up to six flavons; see \Cref{tab:ModelExampleLiterature}.

The next steps are the assignment of irreducible representations to each particle and the construction of the Lagrangian. For instance, in the $A_4$ group, a common choice is to assign the $\SU{2}$ lepton doublets in a triplet irreducible representation~\cite{Ding:2011gt}. 
To construct the Lagrangian, it requires that the invariant contractions under the flavor symmetry be computed, for example, using \texttt{Discrete}~\cite{Holthausen:2011vd} in Mathematica. The generalized superpotential (from which the Lagrangian is extracted) up to dimension five, is then
\begin{align}
\label{eq:ModelsForML}
    \mathcal{W} &~=~ \alpha^{ij}_{(C)} \left( L_i E_j H_d\right) + \frac{\alpha^{ijk}_{(C)}}{\Lambda} \left( L_i E_j  \phi_k  H_d\right)\nonumber\\
&~+ \alpha^{ij}_{(D)}\left( L_i N_j H_u\right) +\frac{\alpha^{ijk}_{(D)}}{\Lambda}\left( L_i N_j  \phi_k H_u\right) \nonumber\\
&~+ \Lambda\alpha^{ij}_{(M)}\left( N_i N_j\right)+\alpha^{ijk}_{(M)}\left( N_i N_j \phi_k\right) + \frac{\alpha^{ijkl}_{(M)}}{\Lambda}\left( N_i N_j \phi_k \phi_l\right)\;,
\end{align}
where the $\alpha$ coefficients are dimensionless constants,$\phi_i$ are the flavons, and $\Lambda$ is both the cut-off scale and the right-handed neutrino mass scale. The first line in \Cref{eq:ModelsForML} is the contribution to the charged lepton mass matrix. The second line is the contribution to the neutrino Dirac mass matrix. Finally, the last line is the contribution to the Majorana mass matrix. The exact form of the superpotential for a given model immediately follows from the particle content and corresponding charge assignments, from which the $\alpha$ parameters can be fit to the experimental observables in Supplementry Tab. 1. 

The flavons are assumed to acquire a \ac{VEV} scale equal to $ 0.1 \Lambda$, and thus influence the overall scale of the Yukawa couplings. On the other hand, the different \ac{VEV} alignment configurations for the flavon triplets determine the mass matrix structure, and hence the neutrino mixing observables. Here, the flavons are chosen to acquire one of the following typical \ac{VEV} configurations~\cite{Altarelli:2005yx,Ding:2011gt}:
\begin{align}\label{eq:vevchoices}
    \braket{\phi} &~=~ 0.1 \Lambda \left(1,0,0\right)\;, \qquad \braket{\phi} ~=~ 0.1 \Lambda \left(1,1,1\right)\;, \qquad \braket{\phi} ~=~ 0.1 \Lambda \left(0,1,0\right)\;, \nonumber\\
    \braket{\phi} &~=~ 0.1 \Lambda \left(0,0,1\right)\;, \qquad \braket{\phi} ~=~ 0.1 \Lambda \left(0,1,-1\right)\;, \qquad \braket{\phi} ~=~ 0.1 \Lambda \left(1,\omega^2,\omega\right) \;,\nonumber \\
    \braket{\phi} &= 0.1\Lambda\left(1, \omega, \omega^2 \right)\;,
\end{align}
where $\omega = \exp\left( \frac{2\pi \mathrm{i}}{3}\right)$. The complex Clebsch–Gordan coefficients (see the supplementary subsubsections “$A_4$ properties” and “$T_{19}$ properties” in the “Supplementary Methods”) along with the complex \acp{VEV} in \cref{eq:vevchoices} are the sources of \CP violation, as the dimensionless coefficients $\alpha$ in \cref{eq:ModelsForML} are assumed to be real. Once the \acp{VEV} are chosen, the mass matrices can be constructed and the values of the effective parameters are obtained numerically through $\chi^2$ minimization using the standard \ac{PMNS} parameterization~\cite{ParticleDataGroup:2024cfk}. 

The number of free parameters, $n_p$, of the model is a strong heuristic estimator of its capacity to generalize to future experiments rather than simply describe current results, and will be a vital metric in the \ac{RL} approach. Dimensionful parameters are factored out by removing an overall scale and expressing related parameters as ratios where possible. For instance, one can always factor out one parameter $\hat{\alpha}_{(C)}$ in the first line in \Cref{eq:ModelsForML}, $\hat{\alpha}_{(D)}$ in the second line and $\hat{\alpha}_{(M)}$ in the third line. The mass scales are then given by $\hat{\alpha}_{(C)}v_d$ for the charged leptons and $\hat{\alpha}_{(D)}v_u^2/\Lambda \hat{\alpha}_{(M)}$ for the active neutrinos. This reduces the number of observables by one for each of the neutrino sectors and the charged lepton sector, and allows the fit to focus on the reduced  parameter space of the mass ratios. Thus, the number of effective parameters is given by the number of dimensionless couplings $\alpha^{(i)}$ minus three. These effective parameters are fit to the observables in Supplementary Tab. 1 with some minimization software. Here, the \texttt{FlavorPy} package~\cite{FlavorPy,Baur:2022hma} is chosen to perform the fits.

\begin{table}[t!]
   \centering
  \caption{\textbf{Example model found by AMBer}. \textbf{(a)} Field assignments under $A_4$ and $\mathbb{Z}_4$ symmetries. The bold number in the $A_4$ column represent the respective irreducible representation for that particle, while the $\mathbb{Z}_4$ charges range from 0 to 3. The model is written using the notation commonly found in the literature. \textbf{(b)} Model representation written as presented to the agent. Non-Abelian symmetries are one-hot encoded while $\mathbb{Z}_4$ charges range from 1 to 4. The last two columns with zero entries indicate to the agent that those two fields are not present.}
  \label{tab:ModelExample}
  \begin{subtable}{0.9\textwidth}
    \centering
    \begin{tabular}{cccccccccccc}
        \toprule
         & $L_i$ & $E_i$ &  $N_i$ & $H_u$ & $H_d$ & $\phi_1$ & $\phi_2$ &   \\
        \midrule
        $A_4$ & $\rep{3}$ & $\rep{3}$ &  $\rep{3}$ & $\rep{1''}$ & $\rep{1}$ & $\rep{3}$ & $\rep{3}$  \\ 
        $\mathbb{Z}_4$ & $2$ & $3$ & $3$ & $3$ & $3$ & $0$ & $3$  \\
        \bottomrule
    \end{tabular}
    \caption{}
    \label{tab:ModelExampleLiterature}
  \end{subtable}

  \begin{subtable}{0.9\textwidth}
    \centering
    \begin{tabular}{lccccccccccccccccc}
        \toprule
        & $L_1$ & $L_2$ & $L_3$ & $E_1$ & $E_2$ & $E_3$ & $N_1$ & $N_2$ & $N_3$ & $H_u$ & $H_d$ & $\phi_1$ & $\phi_2$ & $\phi_3$ & $\phi_4$ & $\phi_5$ \\
        \midrule
        $A_4^{(1)}$   & 0 & 0 & 0 & 0 & 0 & 0 & 0 & 0 & 0 & 0 & 1 & 0 & 0 & 0 & 0 & 0 \\
        $A_4^{(1')}$  & 0 & 0 & 0 & 0 & 0 & 0 & 0 & 0 & 0 & 0 & 0 & 0 & 0 & 0 & 0 & 0 \\
        $A_4^{(1'')}$ & 0 & 0 & 0 & 0 & 0 & 0 & 0 & 0 & 0 & 1 & 0 & 0 & 0 & 0 & 0 & 0 \\
        $A_4^{(3)}$   & 1 & 1 & 1 & 1 & 1 & 1 & 1 & 1 & 1 & 0 & 0 & 1 & 1 &0 & 0 & 0 \\
        $\mathbb{Z}_4$ & 3 & 3 & 3 & 4 & 4 & 4 & 4 & 4 & 4 & 4 & 4 & 1 & 4 & 0 & 0 & 0 \\
        \bottomrule
    \end{tabular}
    \caption{}
    \label{tab:ModelExampleAgent}
  \end{subtable}
\end{table}

A viable neutrino mass model must also satisfy constraints related to the absolute mass scale of neutrinos, namely the effective electron-neutrino mass, $m^{\rm eff}_{\nu_e}$, the effective majorana mass, $m_{ee}$, and the sum of neutrino masses, $\Sigma  m_\nu$. Only dimensionless quantities are used for the fit, so to compare the models found with these dimensionful quantities, an overall active neutrino mass scale $M_\nu$ must be chosen. This scale is chosen to be 
\begin{equation}
    M_\nu^2 = \frac{1}{2} \left( \frac{|\Delta m_{21}^2|_{\rm exp}}{|\Delta \tilde{m}_{21}^2|_{\rm fit}} + \frac{|\Delta m_{31}^2|_{\rm exp}}{|\Delta \tilde{m}_{31}^2|_{\rm fit}}
    \right)\;,
\end{equation}
where the numerator is the experimentally measured value, and the denominator $\Delta \tilde{m}^2_{ij}$ is the dimensionless mass-squared difference that is evaluated from the dimensionless neutrino mass matrix at the fitted parameters. Fits to the $m^{\rm eff}_{\nu_e}, m_{ee},$ or $\Sigma m_\nu$ are not included when searching for models. Instead, these quantities are calculated for filtered models after training. The number of filtered models satisfying bounds set by current experiments will be reported.
Several important aspects of neutrino flavor model-building have not been discussed in full here. For instance, the construction of the flavon potential, lepton flavor violating processes~\cite{Feruglio:2008ht,Feruglio:2009hu}, the \ac{RGE} running of the model parameters ~\cite{Antusch:2005gp}, and the corrections to the K\"{a}hler potential and the superpotential, some of which can induce theoretical uncertainties~\cite{Criado:2018thu,Almumin:2022rml}. 
To start with these, the decay width of lepton flavor violating processes, such as  $\mu\to e\gamma$, has been previously computed in $A_4$ models~\cite{Feruglio:2008ht,Feruglio:2009hu}, where these decays are generated by higher dimensional operators suppressed by a mass scale $M$. The specification of such a mass scale $M$ would render the results sensitive to experimental constraints. However, the mass scale $M$ generally depends on the masses of the heavy particles, which are left unspecified in this work. Furthermore, the value of the charged-lepton mass ratios at the GUT scale obtained from \ac{RGE} is included in Supplementary Tab. 1 assuming only \ac{MSSM} particle content. While it is possible that additional contributions from flavons and right-handed neutrinos could be important, they are generally expected to be negligible if their masses are near the GUT scale. In any case, the neutrino flavor models found here will ultimately require a detailed study of these effects to be considered viable. Such considerations are deferred to future work.

\subsection{Model-Building with RL}
\label{Subsec:MLModelBuilding}
This subsection discusses how neutrino flavor model-building has been cast as an \ac{RL} task, and describes the software optimization, the basics of \ac{RL}, and the specifics of how the \ac{RL} agent searches the space and constructs models. Additionally, an ML-based visualization and diagnostic tool is detailed.

\subsubsection{Software Optimization}
\label{subsubsec:physicsSoftware}
In the model-building recipe outlined in subsection \myref{Subsec:NeutrinoModelBuilding} in the \myref{Sec:Methods} section, several steps can be computationally intensive. Specifically, the calculation of the superpotential, the mass matrices and the parameter fit are bottlenecks that prevent rapid evaluation of a specific model.  To enable an efficient search, an optimized software pipeline is developed to allow automated end-to-end evaluation of a model.

The computation of the superpotential from the particle assignments requires knowledge of the Clebsch-Gordan coefficients of the chosen flavor symmetry group. Existing scientific software such as \texttt{Discrete}~\cite{Holthausen:2011vd} is often used for these calculations. This package extracts the matrix representations of the given group elements through the library GAP~\cite{GAP4}, and computes the Clebsch-Gordan coefficients for the product between irreducible representations~\cite{Broek}. However, the \texttt{Discrete} package is written in Mathematica, which can be slow and cumbersome to integrate into a Python-based \ac{RL} loop. To allow seamless compatibility and computational efficiency within the \ac{RL} environment, \texttt{PyDiscrete}, a Python-based translation of some of the key features of \texttt{Discrete}, was developed. This implementation offers significant speed improvements depending on the task. For example, the Clebsch-Gordan coefficients were computed for all of the different products of three triplets in $T_{19}$ (see Supplementary Tab. 2) and the average computation time was calculated to be $590$ times faster than \texttt{Discrete}.

The symbolic calculation of the lepton mass matrices was automated with a the Python subpackage, \texttt{Model2Mass}. This subpackage uses \texttt{PyDiscrete} and \texttt{SymPy} to construct the superpotential in \cref{eq:ModelsForML} for any finite non-Abelian group, for a given particle content and its respective irreducible representation assignments; see \Cref{tab:ModelExampleLiterature}. It also extracts the charged-lepton mass matrix, the neutrino Dirac mass matrix and the Majorana mass matrix. The \texttt{PyDiscrete} and \texttt{Model2Mass} subpackages are publicly released together in the \texttt{FlavorBuilder} package, installable via the Python Package Index (PyPI), and is available at this \href{https://pypi.org/project/FlavorBuilder/}{link}.

Fitting the free parameters to observables requires minimizing the $\chi^2$ over a large parameter space that features many local minima. Searches for a global minimum often employ heuristics which are computationally expensive and unpredictable. To balance the need to limit computational cost and avoid local minima, \texttt{FlavorPy} is set to perform 100 brief searches from unique initial conditions, with the best resulting $\chi^2$ selected.

\subsubsection{Reinforcement Learning}
\label{subsubsec:ReinforcementLearning}

In \ac{RL}~\cite{10.5555/3312046}, an agent learns a policy to select from available actions to maximize value (roughly, the expected reward) within an environment. Traditional applications include learning policies to make decisions about gameplay~\cite{Silver2017} or navigation~\cite{INAMDAR2024100810}. \ac{RL} has also been applied to search theoretical physics spaces, including string theory landscapes~\cite{Halverson_2019}, Froggatt-Nielsen models of quark and lepton masses~\cite{Nishimura:2020nre, Harvey:2021oue, Nishimura:2024apb}, dark matter models with vector-like leptons~\cite{Wojcik:2024lfy}, and supersymmetric gauge theories~\cite{Carta:2025asr}.  In
previous applications, an agent has been able to find regions of quality models within large theory spaces without requiring
an exhaustive search. These approaches have sought to limit the need to interface the agent with complex scientific software,
either by fixing the available symmetries and couplings ~\cite{Nishimura:2020nre, Harvey:2021oue, Nishimura:2024apb}, using embeddings strategies~\cite{Wojcik:2024lfy}, or avoiding the statistical fits. However, it is essential to overcome these challenges
and interface the agent with scientific software in order to extend this approach to larger theoretical spaces. The approach taken here is to design a framework for integrating existing scientific software with RL, in a similar spirit to previous efforts in interfacing physics software with probabilistic programming tools ~\cite{Baydin_2019}.

Here, the search for a neutrino flavor model which matches observation with few free parameters is cast as an \ac{RL} task. The agent's environment can be considered the space of possible models, with its state the current model being evaluated. The available actions include changes to the particle content, changes to the particle representations, or in some cases changes to the symmetries imposed. In addition to fixed intervals for model evaluation, the agent can choose to evaluate the model at intermittent steps. The value function is the agent's expected return, which is the discounted sum of future rewards, with more distant rewards weighted less than near-term rewards. The reward for a model is generally parameterized by its $\chi^2$ and number of parameters. The agent's goal is therefore to maximize its expected return, encouraging it to take actions that ultimately produce high-quality models even if the immediate reward is low. The training occurs over a series of episodes where each episode starts with an initial state and continues either until a maximum number of steps are taken, or a terminal state is reached. A terminal state is one for which the reward exceeds a certain threshold. When a terminal state is achieved, the agent is given a large reward and is reset to a random point in the theory space, starting a new episode.

 \ac{RL} algorithms develop a policy to select actions, a task often enhanced by simultaneously learning to predict the expected return given the current state. \ac{PPO} \cite{DBLP:journals/corr/SchulmanWDRK17} optimizes two neural networks, one parameterizing the policy and one for predicting the expected return. During training, the agent utilizes the policy network to select actions, and receives rewards based on the new state produced by each action. Periodically, the value network and policy network are updated in order to maximize the accuracy of the value network's predictions and the advantage of the actions chosen by the policy network. The networks utilize a loss function composed of three terms: 
\begin{enumerate}
    \item  Value loss, the mean squared error between the predicted return and the actual return.
    \item  Policy loss, the advantage of action taken over average of all possible actions, weighted by ratio of the probability of given action from old and new policies. The policy loss is clipped to prevent excessively large updates.
    \item Entropy loss, the Shannon entropy of the policy, encourages exploration of a broad distribution across the action space.
\end{enumerate}

Although both networks share the loss function, the value loss is a function only of the value network parameters, while the policy and entropy losses are functions of only the policy network parameters. With this method, the value network learns to improve its predictions of the expected return, and the policy network learns to determine which actions will lead to the highest future return~\cite{DBLP:journals/corr/SchulmanWDRK17}. While one could experiment with alternative RL algorithms, PPO implemented with \texttt{Stable-Baselines3}~\cite{stable-baselines3} worked sufficiently well for this task.

\subsubsection{Autonomous Model Builder}
\label{subsubsec:AutonomousModelBuilder}

The Autonomous Model Builder (AMBer) described here is an \ac{RL} agent with the task to learn a policy that will change a neutrino flavor model to make it more accurate while minimizing the number of free parameters, beginning from a random initial model. 

\paragraph*{Environment}\mbox{}\\
The model being considered is the input to the policy and value networks. Each particle in a model is represented by a one-hot encoded vector to indicate its representation under the non-Abelian symmetry, with an additional element indicating the particle's charge under the Abelian symmetry. Several rules are incorporated into the environment to ensure that changes to the model are mathematically consistent. These include having a representation for all leptons and Higgses (not required for flavons), consistent representation of particles (for example, if one right-handed charged lepton is in a triplet all three must be), and consistent Abelian charges (i.e., all members of a triplet must have the same charge). The environment's current state is represented as a matrix; see~\Cref{tab:ModelExampleAgent} for an example. 

Additional components of the model provide important context, including a vector representing which leptons are associated together in a triplet, a vector representing the \ac{VEV} configuration for each flavon, and a vector indicating the order of the present Abelian symmetry (i.e. the number of elements in the group). 

Finally, the agent only evaluates the reward every $n^{th}$ step, so the number of steps until the next reward evaluation is also included in the environment. At each step, \ac{AMBer} observes the entire state of the environment before selecting an action.

\paragraph*{Actions}\mbox{}\\
The agent can change the representation of a particle (or multiplet), its charge under an Abelian symmetry, the \ac{VEV} configuration of a flavon triplet, the number of flavons, and in some cases the order of the Abelian symmetry. Each representation of the non-Abelian group can be changed in one action from any other representation. For the $\mathbb{Z}_N$ charge, two types of actions are defined. The first shifts the $\mathbb{Z}_N$ charge of a single field by $\pm 1$. For the second action, all fields, excluding the flavon fields, with the same \ac{SM} representation have their $\mathbb{Z}_N$ charge changed to $\pm 1$ of the first particle (e.g., $L^1$, $E^1$, or $N^1$). Under this choice, constructing mass matrices of rank $3$ becomes easier than achieving ranks of $2$ or $1$. Finally, the agent can choose to evaluate the model between its set evaluation steps in order to gain more information on recent actions. Dynamic decisions to run physics software are a challenge for traditional search algorithms, but can be naturally incorporated in RL. All possible actions for each sector are listed in Supplementary Tab. 3.

\paragraph*{Reward}\mbox{}\\
The reward function $R$ for \ac{AMBer} is defined as
\begin{align}
    R(\chi^2, n_p, N) := \begin{cases}
        -c_{\mathrm{inv}}\;,\;\;\qquad\qquad\qquad\qquad\qquad\qquad\mathrm{if\;invalid\;action}\\
        -c_{\mathrm{rank}}-\frac{9-\Sigma_i \mathrm{rank}(M_i)}{9}\;,\qquad\qquad\quad\;\;\;\;\;\;\mathrm{if\;invalid\;model}\\
        c_1R_\chi(\chi^2) + c_2R_p(n_p) + c_3R_{\mathbb{Z}}(N),\;\;\;\;\quad\mathrm{otherwise}\\
         \end{cases}
    \label{eq:reward}
\end{align}
where $c_{\mathrm{rank}},c_{\mathrm{inv}}>0$ quantify the penalty (i.e. a negative reward) for an invalid action or model respectively, $c_{j}$, for $j=1,2,3$, are  positive coefficients and $N$ is the order of the Abelian group $\mathbb{Z}_N$. 

Invalid actions include those that do not change the model, to avoid stagnation, or actions which change the Abelian charge of an absent flavon. These invalid actions must be learned in this implementation as the action space is fixed and cannot dynamically exclude invalid actions. Invalid models are those that are obviously in conflict with experiment, such as predicting massless leptons, as in the case where $M_C$ has a rank less than $3$. Another possibility of invalid model is the case with $2$ or more massless neutrinos. Assuming three right-handed neutrinos, this requires a valid model to have a rank $\geq 2$ for $M_D$ and rank $3$ for $M_M$. Since this is known prior to performing a fit, the invalid model penalty allows the agent to avoid unnecessary computations. The invalid model penalty includes a term associated with the sum of the rank of the three mass matrices, which encourages the agent to increase the rank in order to reduce the penalty, leading the agent to consider models with more mass terms.

In case of a valid action and model, the coefficients $c_{j}$ scale the relative importance of the terms

\begin{align}
    &R_\chi(\chi^2) := -c_{scale}\tanh{[\frac{\log_{10}(\chi^2 - \chi^2_{\mathrm{target}})}{\log_{10}(\chi^2_{\mathrm{target}})}-1]}-c_{shift} \;, \label{eq:rewardChi}\\
    &R_p(n_p) := \log_{10}\frac{n_p^{\mathrm{target}}}{n_p}\;, \label{eq:rewardP}\\
    &R_{\mathbb{Z}}(N) := \frac{(N - N_{\mathrm{min}} - N_{\mathrm{target}})^3}{N_{\mathrm{max}} - N_{\mathrm{min}}}\;. 
\label{eq:rewardZ}
\end{align}

The $R_\chi$ term  encourages the agent to minimize the $\chi^2$ fit. The $\chi^2_{\mathrm{target}}$ is a parameter which varies over the run to modify the strictness of the $R_\chi$ term, while $c_{scale}$ and  $c_{shift}$ are constants which set the range of the reward and are fixed over the run. The $R_p$ term encourages the agent to minimize the number of parameters in the model, $n_p$. The parameter $n_p^{\mathrm{target}}$ shifts over the course of the run to control the strictness of this term. The $R_\mathbb{Z}$ term in \cref{eq:rewardZ} encourages exploration by including a reward for higher order models. This term is used only when the agent is allowed to alter the order of the Abelian symmetry. In these circumstances, the agent can choose $N=2,\dots,10$, so $N_{\mathrm{min}} = 2$ and $N_{\mathrm{max}}=10$ are fixed. $N_{\mathrm{target}}$ determines the degree of preference towards higher order Abelian groups. 

On the steps where the agent does not evaluate the model, it receives a reward of 0. If it selects the action to evaluate the model on an intermittent step, the agent receives a reward based on whether it has improved the model since its last evaluation. If the reward has improved, instead of receiving the actual reward for that model the agent is given a reward just below the threshold. If the reward does not improve from its last value, the agent receives a reward equal to the invalid model case, $-c_{rank}$. This encourages the agent to only use this intermittent evaluation action when it thinks the model has improved.

Finally, the reward is normalized to maintain values between -1 and 1.
When the reward exceeds 0.1, it is boosted to 1 to emphasize that state. For more details on the behavior of the reward function, see the “Reward Function” subsubsection in the “Supplementary Methods”.

\paragraph*{Training}\mbox{}\\
\ac{AMBer} trains with 250 parallel environments to maximize CPU utilization. Each environment has a maximum episode length of 1000 steps a full run continues until 32000 steps are taken in each environment. The neural networks of the agent use the default \texttt{Stable-Baselines3} PPO hyperparameters, with the exception of the entropy coefficient, which is set to 0.05. The networks have 2 fully connected hidden layers with 64 nodes each, and are randomly initialized, as are the initial state models. Training occurs every 32 steps through the environments. In addition to when the agent chooses to evaluate, models are evaluated every $n$ steps, where $n$ is a tunable parameter. This allows several actions to be taken without any reward or penalty, which permits the agent to pass through poor models without reducing its reward. The $A_4$ runs use $n=10$, while the $T_{19}$ runs use $n=5$. 

The performance goals, $\chi^2_{\mathrm{target}}$, $n_p^{\mathrm{target}}$ and $N_{target}$, shift over the training, encouraging the agent to build upon lessons it learns early in the run in order to find increasingly high quality models. The values of these targets are detailed in Supplementary Tab. 4. The reward function is plotted in Supplementary Fig. 2.  
Each training run consists of multiple independent agents with identical hyperparameters and environments in order to identify general trends. Training each agent takes approximately 10 hours running on a single dedicated CPU. Here, training runs are performed in three theory spaces:
\begin{enumerate}
    \item $A_4\times \mathbb{Z}_4$ with $n_{\phi}\leq 5$, where $n_{\phi}$ is the number of flavons.
    \item $A_4 \times \mathbb{Z}_N$ where the agent can alter $N$ in the range $[2, 10]$ and $n_{\phi}\leq 5\;.$
    \item $T_{19}\times\mathbb{Z}_4$ with $n_{\phi}\leq 6\;.$
\end{enumerate}
The upper bound on the number of flavons $n_{\phi}$ is motivated by models in the literature~\cite{Ding:2011gt}. For $T_{19}$, $n_\phi\leq 6$ is chosen such that \ac{AMBer} has the ability to place one flavon in each triplet representation, if it so chooses. All hyperparameters are manually tuned.

\begin{figure}[h!]

    \centering
    \includegraphics[width=1.0\textwidth]{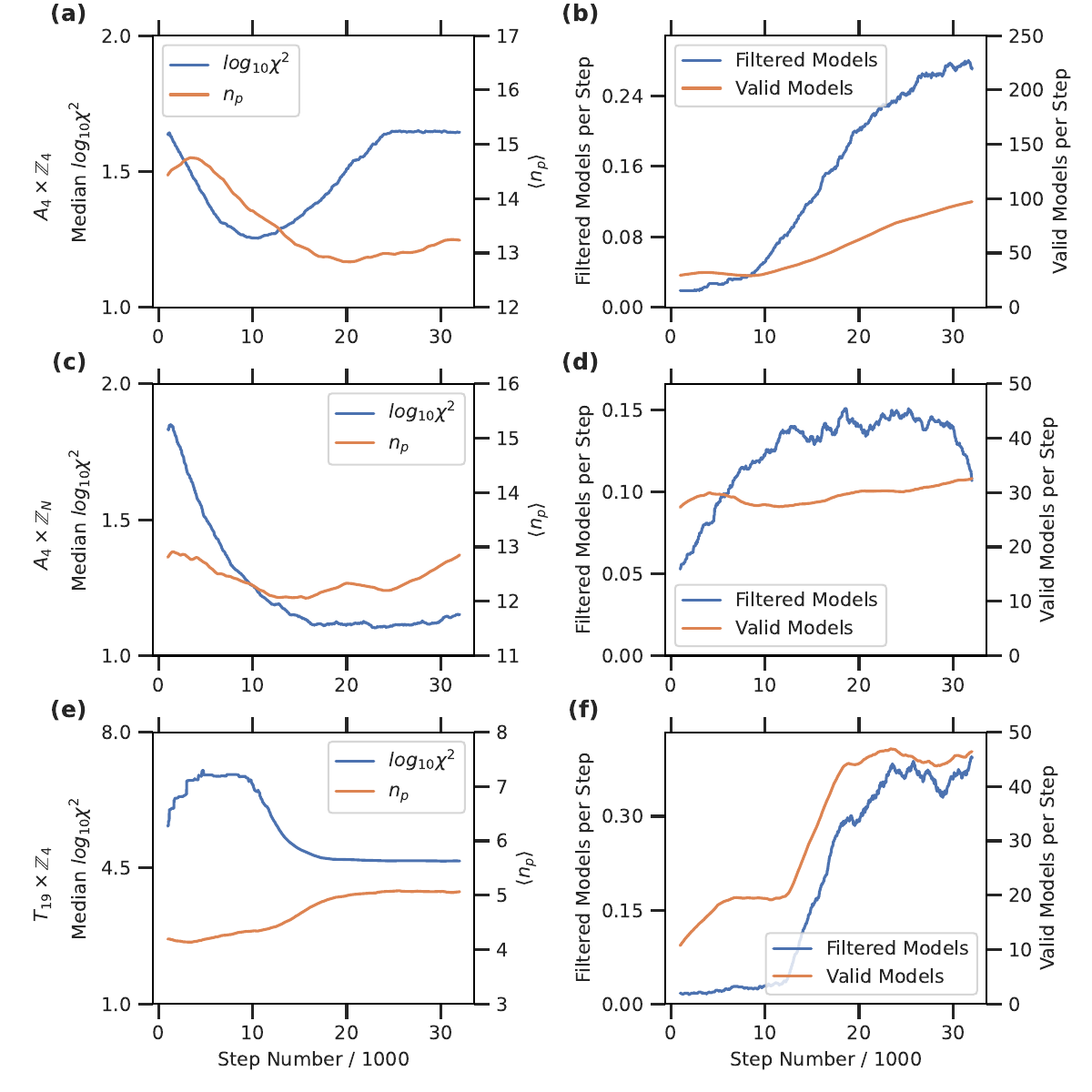}

    \caption{\textbf{Evolution of variables of interest over training.} Searches are performed in three spaces: $A_4 \times  \mathbb{Z}_4$ (panels \textbf{(a)} and \textbf{(b)}), $A_4 \times \mathbb{Z}_N$ (panels \textbf{(c)} and \textbf{(d)}), and $T_{19} \times \mathbb{Z}_4$ (panels \textbf{(e)} and \textbf{(f)}). Panels \textbf{(a)}, \textbf{(c)} and \textbf{(e)} show the evolution of $\chi^2$ in blue (where the curve indicates the median $\log_{10}{\chi^2}$ over all environments) and the mean number of parameters $\langle n_p \rangle$ as training progresses in orange for the different theory spaces. Panels \textbf{(b)}, \textbf{(d)} and \textbf{(f)} show the number of valid models in orange and filtered ($\chi^2 \leq 10$ and $n_p \leq$  7) models in blue for the three theory spaces. These are smoothed to more clearly see the relevant trends. There are two y-axes for each panel due to the differences in scale of each variable. Supplementary Tab. 4 describes how the reward targets change over time.}
    \label{fig:RewardDiagnostics}    
\end{figure}

\paragraph*{Diagnostics for \ac{AMBer} runs}\mbox{}\\
Performance during training can be evaluated by tracking the $\chi^2$ and $n_p$ metrics, the number of valid actions and models, and network-level quantities.

The left column of \Cref{fig:RewardDiagnostics} shows how the $\chi^2$ and number of parameters $n_p$ evolve over the course of a run. In panel \textbf{(a)}, corresponding to an $A_4 \times \mathbb{Z}_4$ run, the average number of parameters per model decreases over time, while the median $\chi^2$ initially decreases before returning to a value of $\sim 50$. This demonstrates the tradeoff between the competing goals of simplicity and fitness. In panel \textbf{(c)}, corresponding to an $A_4 \times \mathbb{Z}_N$ run, the average number of parameters changes little, while the $\chi^2$ steadily decreases. In panel \textbf{(e)}, corresponding to a $T_{19} \times \mathbb{Z}_4$ run, there is a more complex trade-off between the two training goals. The average number of parameters starts much lower in this space, meaning the agent must learn to increase the number of parameters. As it does this, the median $\chi^2$ decreases as it becomes easier to find a good fit with more parameters.

The right column demonstrates how the number of valid models and filtered models (those with $\chi^2 \leq 10, n_{p} \leq 7$) evolve over time. In all three theory spaces, the agent learns to build valid models over time due to the rank penalty, and increases its rate of filtered models due to the $R_\chi$ and $R_p$ terms in the reward. In the $A_4 \times \mathbb{Z}_N$ run shown, the increase in valid models per step is more modest as the agent learns to build invariant terms with each different Abelian symmetry. Additionally, the number of filtered models per step decreases at the end, likely indicating the agent has become influenced too heavily by one reward term over the others. This hypothesis is supported by the late increase in the number of parameters in panel \textbf{(c)}.

\Cref{tab:RandomComp} details the number of filtered models found in random scans versus AMBer training runs. In the $A_4 \times \mathbb{Z}_4$ and $A_4 \times \mathbb{Z}_N$ searches, 5 independent agents are trained, while in the $T_{19} \times \mathbb{Z}_4$ run 20 independent agents are trained in order to capture greater run-to-run variation. Since $T_{19}$ has more representations, and therefore more plausible permutations of those representations, it is more susceptible to mode collapse, over-preferring a particular permutation. Additionally included are the number of inequivalent models and the number of models that satisfy dimensionful experimental constraints. A model is considered inequivalent if it does not share with any other model identical analytical expressions for their mass matrices, and if their irreducible representations under the flavor groups differ, even after accounting for possible field permutations. For instance, exchanging the irreducible representations of the $E_1$ and $E_2$ fields yields the same physical model, since this operation merely permutes the columns of the charged lepton mass matrix. The same reasoning applies to permutations among the $L_i$, $N_i$, and flavon fields $\phi_i$ (taking into account their respective \acp{VEV}). Hence, the numbers quoted in \Cref{tab:RandomComp} indicate that \ac{AMBer} is not settling into a specific set of neutrino flavor models given our definition of inequivalent models. A fully exhaustive analysis would require checking whether any two models are related by a unitary transformation.

For the filtered models, the observables related to the absolute mass scale of neutrinos, $m^{\rm eff}_{\nu_e}$, $m_{ee}$ and $\Sigma m_\nu$, are calculated. The experimental constraints on these are given by $m^{\rm eff}_{\nu_e}<0.45~{\rm eV}$ from KATRIN ~\cite{KATRIN:2024cdt}, $m_{ee}<36~{\rm meV}$ from KamLAND-ZEN~\cite{KamLAND-Zen:2022tow} and $\Sigma m_\nu<0.12~{\rm eV}$ from Planck~\cite{Planck:2018vyg}.  The number of inequivalent filtered models that satisfy these constraints is reported in \Cref{tab:RandomComp}.

AMBer outperforms a random scan in all spaces. To verify this efficiency gain, the number of CPU-hours for a random scan and a training run must be compared. On average, a training run takes approximately 70 percent longer to complete than a random scan. This can be attributed to the fact that \ac{AMBer} identifies many more valid models, requiring more expensive computations than the random scan. Nonetheless, the number of filtered models per CPU-hour is still significantly higher for AMBer than for a random scan. Additional neural network level diagnostics are shown in Supplementary Fig. 3.

\begin{table}[t!]
\centering
\caption{\textbf{Performance comparison of \ac{AMBer} and random scans.} Each method evaluates the same number of models; results are shown separately for \ac{NO} and \ac{IO}. Reported as $x/y/z$: total number of filtered models, the number of inequivalent filtered models, and the number of inequivalent filtered models satisfying constraints from beta decay, neutrinoless double beta decay, and cosmological neutrino mass limits. The fewer inequivalent filtered models satisfying these constraints for inverted versus normal ordering reflects the generally heavier sum of neutrino masses in IO, which make the bounds more restrictive.}
\label{tab:RandomComp}
\begin{tabular}{c c c c c c} 
  \toprule
    Space & Models Seen  & Random scan NO & Random Scan IO &  \ac{AMBer} NO &  \ac{AMBer} IO \\ \midrule
    $A_4 \times \mathbb{Z}_4$ & $4 \times 10^6$ & 28/28/16 & 20/20/3 & 733/683/402 & 518/463/26\\
    $A_4 \times \mathbb{Z}_N$ & $4 \times 10^6$ & 21/21/19 & 21/21/6 & 1404/1394/837 & 1179/1170/87\\ 
    $T_{19} \times \mathbb{Z}_4$ & $24 \times 10^6$ & 409/409/309 & 244/244/44 & 6471/6439/4535 & 8678/8291/688\\
    \bottomrule
\end{tabular}
\end{table}

\section{Results}
\label{sec:results}

\begin{figure}[h!]
    \centering
    \includegraphics[width = \textwidth]{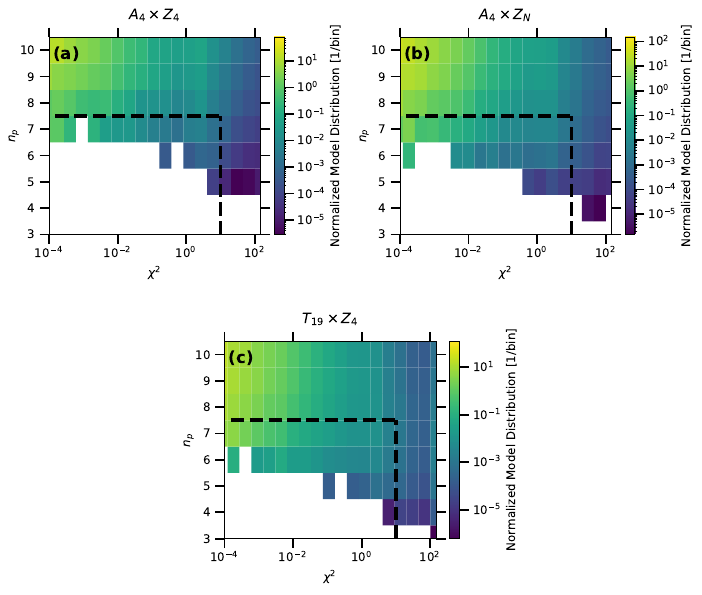}

    \caption{\textbf{Number of parameters $n_p$ and $\chi^2$ for a representative distribution of found models}. Panel \textbf{(a)} shows the distribution for $A_4 \times \mathbb{Z}_{4}$, panel \textbf{(b)} for $A_4\times \mathbb{Z}_{N}$, and panel \textbf{(c)} for $T_{19}\times \mathbb{Z}_{4}$. The region within the dashed black lines contains models with  $\leq 7$  parameters, and good fits, $\chi^2\leq10$.}
    \label{fig:2d_summary}
\end{figure}

This section describes the models found in each theory space. \ac{AMBer} finds too many models to describe each one in detail, so broad features are highlighted, and selected individual models are analyzed in further detail. The complete list of inequivalent filtered models found by \ac{AMBer} is available as a \texttt{Github} repository (\href{https://github.com/jake-rudolph-1/models-by-AMBer}{link}) and a model from each search is presented in the “Supplementary Results”. The three theory spaces outlined in subsection "Model-Building with RL" are considered. In~\Cref{fig:2d_summary}, the distribution of the number of parameters and $\chi^2$ values is shown for a representative sample of models.  \ac{AMBer} finds many models which are predictive ($n_p \leq 7$) and  well fit to data ($\chi^2 \leq 10$), which is denoted by the region bounded by the dashed black line in all three theory spaces. Here, models that satisfy the latter requirements are referred to as “filtered models”. As discussed at the end of subsection "Neutrino Flavor Model-Building", there are several different aspects of these models requiring further study, however models will be analyzed here using only these two conditions. The results for each individual theory space are now discussed.

\subsection{\texorpdfstring{Search in $A_4\times \mathbb{Z}_{4}$ with $n_{\phi}\leq 5$}{Search in A4 × Z4 with nphi ≤ 5}}
\label{subsec:A4XZ4}

\begin{figure}[t]
    \centering
    \includegraphics[width = 0.99 \textwidth]{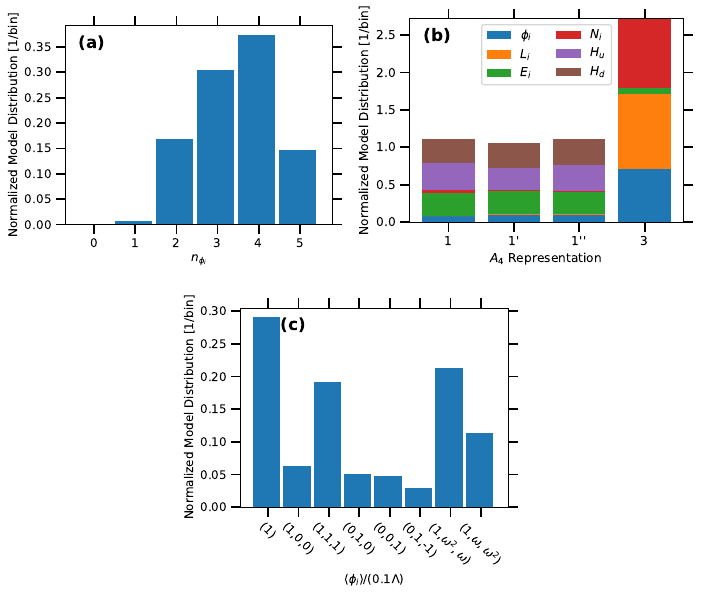}
    \caption{\textbf{$A_4\times \mathbb{Z}_4$ model space distributions of filtered models.} Panel \textbf{(a)} shows the distribution of the number of flavon fields, panel \textbf{(b)} the $A_4$ representations distribution, and panel \textbf{(c)} the vacuum alignments distribution. The vacuum alignment is normalized by the flavon breaking scale, $0.1 \Lambda$, i.e. $\langle\phi \rangle/(0.1\Lambda) $ as in \cref{eq:vevchoices}.}
    \label{fig:a4_z4_flavondistribution}
\end{figure}

In the $A_4\times \mathbb{Z}_{4}$ space, \ac{AMBer} is allowed up to five flavon multiplets. More flavons allow extra flexibility in the model by producing additional mass terms, but at the cost of introducing additional parameters.  \ac{AMBer} finds many filtered models (see \Cref{tab:RandomComp}), though it is important to note that some may have mass matrices related by unitary transformations which would make them equivalent.

\Cref{fig:a4_z4_flavondistribution} shows the distribution of the number of flavons,  $n_{\phi_i}$, the $A_4$ representations, and the flavon vacuum alignments of the filtered models. The Higgs doublets are only allowed to reside in singlet representations of $A_4$. Nearly all of the found models find the $L_i$ and $N_i$ fields packed into their respective triplets, as opposed to $E_i$ which is found to be evenly distributed across the singlet representations. This assignment has frequently been explored in the literature~\cite{Ma:2001dn,Ding:2011gt,Altarelli:2005yx}, and provides a good fit to the data with a small number of parameters. \ac{AMBer}'s arrival at this representation distribution indicates its capacity to reproduce reasonable results in familiar settings. The lower panel of \Cref{fig:a4_z4_flavondistribution} shows the distribution of flavon vacuum alignments normalized to the breaking scale, $\langle \vec{\phi} \rangle/(0.1\Lambda) $. Among the triplet representations, the configurations that do not have a zero value in them are preferred. In ~\Cref{tab:ModelExample}, a particular model from this search space is shown that has $\chi^2=5.6$ and 5 parameters.

\subsection{\texorpdfstring{Search in $A_4\times \mathbb{Z}_{N}$ with $n_{\phi}\leq 5$}{Search in A4 × ZN with nphi ≤ 5}}
\label{subsec:A4xZN}

\begin{figure}[t!]
    \centering
    \includegraphics[width=1.0\linewidth]{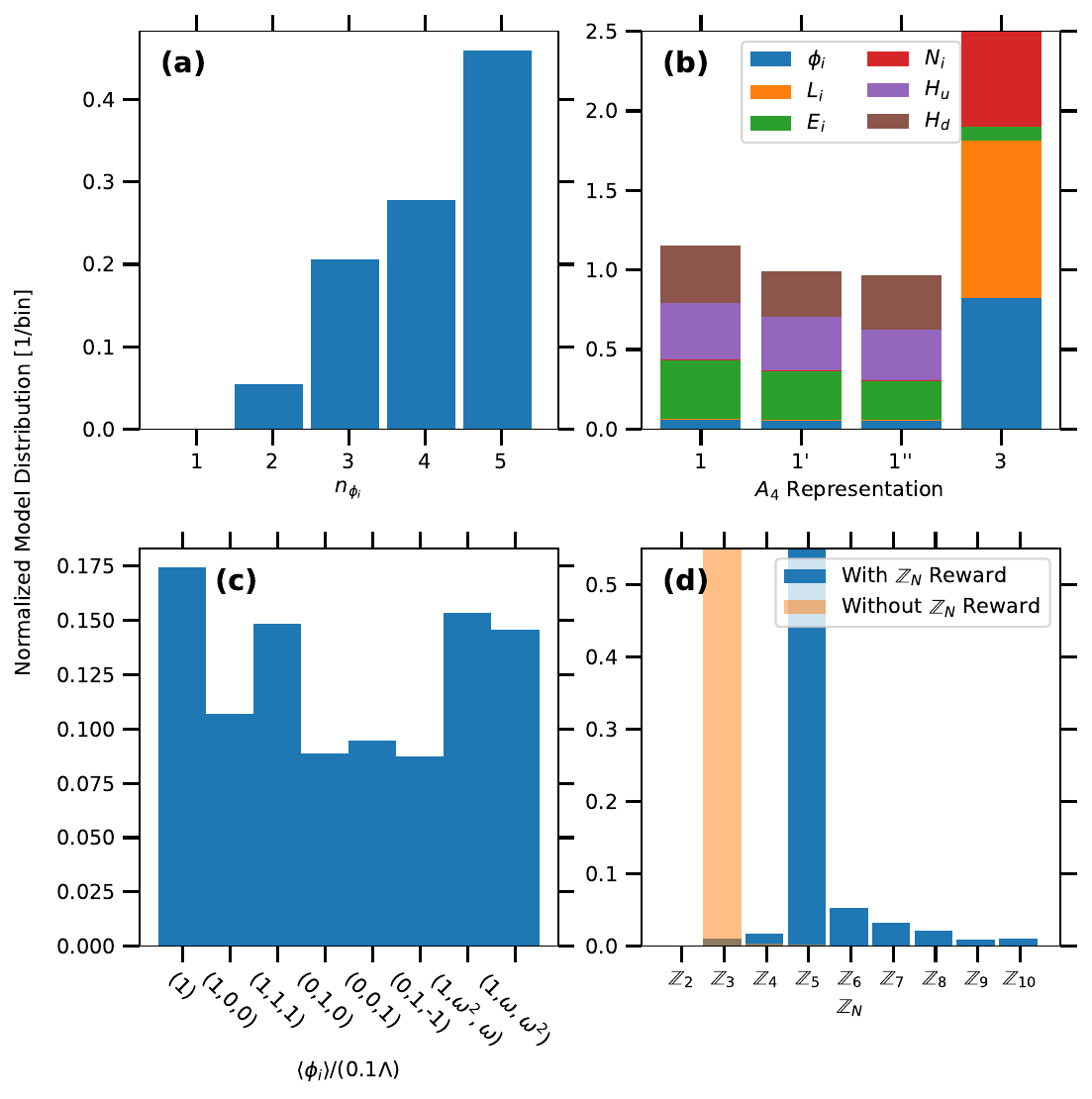}
    \caption{\textbf{$A_4\times \mathbb{Z}_N$ model space distributions of filtered models}. Panel \textbf{(a)} shows the distribution of the number of flavon fields, panel \textbf{(b)} the $A_4$ representations, panel \textbf{(c)} the distribution of vacuum alignments  and panel \textbf{(d)} the $\mathbb{Z}_{N}$ symmetry. For comparison, panel \textbf{(d)} shows a configuration with and without the $R_{\mathbb{Z}}(N)$ reward in \cref{eq:rewardZ}, and \ac{AMBer} is found to be more efficient at exploring large $N$ with this reward included. The distribution fraction for $\mathbb{Z}_3$ models is $\sim0.9$ without the $R_{\mathbb{Z}}(N)$ reward term, but is cut off to better display the distribution of higher order models. The vacuum alignment is normalized by the flavon breaking scale, $0.1 \Lambda$, i.e. $\langle \vec{\phi} \rangle/(0.1\Lambda) $ as in \cref{eq:vevchoices}.}
    \label{fig:a4_zN_distribution}
\end{figure}

In this theory space, \ac{AMBer} has more flexibility to select the imposed symmetry. \Cref{fig:a4_zN_distribution} shows the distribution of the number of flavon multiplets  $n_{\phi_i}$, the $A_4$ representations, the flavon vacuum alignments of the filtered models and the dimension of the Abelian symmetry $\mathbb{Z}_N$.  \ac{AMBer} chooses the number of flavons by balancing the goodness of fit with the number of parameters. Similarly to subsection \myref{subsec:A4XZ4} in the \myref{sec:results} section, \ac{AMBer} typically places the $L_i$ in triplets, and a similar distribution of vacuum expectation values is found.

The bottom right panel shows the distribution of the $\mathbb{Z}_{N}$ symmetry of the found models, which \ac{AMBer} is free to select in this search. With $c_3=0.35$ enabling the $R_{\mathbb{Z}}(N)$ reward term (see \cref{eq:reward}) and $N_{\mathrm{target}}$ evolving with the schedule shown in Supplementary Tab. 4, \ac{AMBer} is encouraged to explore larger $N$. While this choice results in a large number of models with $N=5$, it also increases the number of models with $N>3$ in general. For comparison, this panel also shows the distribution of models found for a separate run that did not have the $R_{\mathbb{Z}}(N)$ in the reward function and \ac{AMBer} is found to focus on $\mathbb{Z}_3$. $\mathbb{Z}_2$ is very rarely employed. This is likely due to the fact that for $\mathbb{Z}_2$, it would be difficult to accommodate a well-fit model with a small number of parameters.

\subsection{\texorpdfstring{Search in $T_{19}\times \mathbb{Z}_{4}$ with $n_\phi\leq6$}{Search in T19 × Z4 with nphi ≤ 6}}
\label{subsec:T19xZ4}

In the $T_{19}\times \mathbb{Z}_{4}$ space with up to six flavon multiplets, \ac{AMBer} explores a subspace of models that has not been explored in the literature.  This search space is more complicated due to the larger $T_{19}$ group. In this group, there are six irreducible triplet representations which can be multiplied to form a number of singlet or triplet representations, as shown in Supplementary Tab. 2. \Cref{fig:t19_distrubtion} shows the distribution of found models for this search. As compared to the $A_4$ run, a more uniform distribution of models across the different representations are found, although there is a slight preference to place $E_i$ in singlets, $L_i,N_i$ in respective triplets

This flavor symmetry allows \ac{AMBer} to pack all lepton flavors, $L_i,E_i,N_i$, into their respective triplets, and still obtain the observed hierarchy. For example, consider the Dirac mass term for the charged-lepton sector, $\bar{L}E\phi_i H_d$ where $L$ and $E$ are in the $3_1$ and $3_2$ representation, respectively. Their tensor product decomposes into a $3_2 \oplus 3_3 \oplus \bar{3}_3$, which can subsequently be made into a singlet if there is a flavon in any of the conjugate representations present in the decomposition. This also points to a certain flexibility in this search space, as there are generally many ways to form group invariant interactions, particularly for Majorana masses of the right-handed neutrinos. In order to reduce parameters, the $\mathbb{Z}_4$ charges are likely being tuned to reduce the free parameters for models.

\begin{figure}[t!]
    \includegraphics[width = \textwidth]{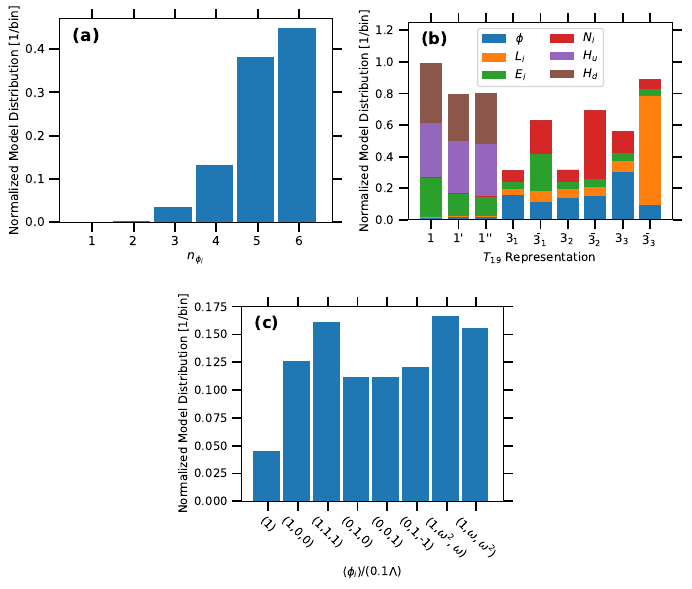}

    \caption{\textbf{$T_{19}\times \mathbb{Z}_4$ model space distributions of filtered models}. Panel \textbf{(a)} shows the distribution of the number of flavon fields, panel \textbf{(b)} the $T_{19}$ representations, and panel \textbf{(c)} the distribution of vacuum alignments. The vacuum alignment is normalized by the flavon breaking scale, $0.1 \Lambda$, i.e. $\langle \vec{\phi} \rangle/(0.1\Lambda) $ as in \cref{eq:vevchoices}.}
    \label{fig:t19_distrubtion}
\end{figure}

\begin{table}[H]
\centering
 \caption{\textbf{Example model found by AMBer in the $T_{19}\times Z_4$ search}. It is discussed in further detail in the main text. Here, $\langle\phi_1\rangle/(0.1\Lambda)=(1,\omega^2,\omega)$,$\langle\phi_2\rangle/(0.1\Lambda)=(1,\omega,\omega^2)$, and $\langle\phi_3\rangle/(0.1\Lambda)=(0,0,1)$.}\label{tab:foundt19model}
\begin{tabular}{llllllllllllll}
\toprule
  & $L$  & $E_1$ & $E_2$ & $E_3$ & $N$ & $H_u$ & $H_d$ & $\phi_1$ & $\phi_2$ & $\phi_3$ \\ 
 \midrule
  $T_{19}$  & $\rep{\bar{3}}_1$ & $\rep{1'} $  & $\rep{1} $  & $\rep{1'}$ & $\rep{3}_3$  & $\rep{1''}$  & $\rep{1'}$ & $\rep{3}_1$ &$\rep{3}_1$  & $\rep{3}_3$  \\ 
 $Z_4$&   $0$ & $1$ & $3$ & $3$ & $0$ & $2$ & $1$ & $0$ & $2$ & $2$  \\ \bottomrule
\end{tabular}

\end{table}

A particularly simple model is shown in \Cref{tab:foundt19model}. The superpotential is given by
\begin{align}
\label{eq:ExampleSuperpotential}
    \mathcal{W} &~=~ \hat{\alpha }_{(C)}\left[\left(L \,E_1\,H_d\,\phi_2 \right)+\alpha_1\left(L \,E_3\,H_d\,\phi_1 \right) + \alpha_2 \left( L E_2 H_d \phi_1 \right)\right]+  \hat{\alpha }_{(M)}\left[\left( N\,N\,\phi_1\,\phi_1\right) + \alpha_3\left( N\,N\,\phi_2\,\phi_2\right)\right]\\ \nonumber
    &~~+\hat{\alpha }_{(D)}\left[\left( N\,L\,H_u\,\phi_2\right) + \alpha_4 \left( N\,L\,H_u\,\phi_3\right)\right]\;.
\end{align}
The best-fit point in the parameter space of this model is at the values given by
\begin{equation}
\begin{aligned}
\alpha_1 &= 0.000271\;, & \alpha_2 &= 0.0565\; , \quad& \alpha_3 &= -0.343\;,  & \alpha_4 &= 0.915\; .
\end{aligned}
\label{eq:BestFitFlavorObservables}
\end{equation}
Hence, with only 4 free parameters it yields $\chi^2\approx 10$.
This particular model places the charged lepton doublets and the right-handed neutrinos in separate triplets, and the right-handed charged leptons in singlets---an alternative model is shown in Supplementary Tab. 7 that has them all in triplets and has comparable performance. After the flavons acquire a \ac{VEV}, the mass matrices can be written as

\begin{equation*}
    m_D = v_u \hat{\alpha}_{(D)}\begin{bmatrix}
0 & -\frac{1}{2} - i\,\frac{\sqrt{3}}{2} & \alpha_4 \\[6pt]
0 & 0 & -\frac{1}{2} + i\,\frac{\sqrt{3}}{2} \\[6pt]
1 & 0 & 0
    \end{bmatrix}\;,
\end{equation*}
\begin{equation}
\label{eq:ExampleModelT19_model}
    m_C = v_d~\hat{\alpha }_{(C)}
    \begin{bmatrix}
-1/2 + i\sqrt{3}/2 & -\alpha_2\,(1/2 + i\sqrt{3}/2) & -\alpha_1\,(1/2 + i\sqrt{3}/2) \\[6pt]
1 & \alpha_2 & \sqrt{3}\,\alpha_1\,(-\sqrt{3}/6 - i/2) \\[6pt]
(0.5 - i\sqrt{3}/2)^2 & \sqrt{3}\,\alpha_2\,(-\sqrt{3}/6 + i/2) & \alpha_1\,(1/2 - i\sqrt{3}/2)^2
    \end{bmatrix},~~ 
\end{equation}
\begin{equation*}
    m_M = \Lambda \hat{\alpha}_{(M)}
    \begin{bmatrix}
(-\alpha_3-1)+i\sqrt{3}(\alpha_3-1) & 0 & 0 \\[6pt]
0 & 2(\alpha_3 + 1) & 0 \\[6pt]
0 & 0 & (-\alpha_3-1)+i\sqrt{3}(-\alpha_3+1)
    \end{bmatrix},
\end{equation*}
where $\hat{\alpha}_{(C)}$, $\hat{\alpha}_{(M)}$ and $\hat{\alpha}_{(D)}$ are constants that can be factored out, because only dimensionless mass ratios are fit (see discussion at the end of \Cref{Subsec:NeutrinoModelBuilding}). The predicted values for the lepton observables are given by
\begin{equation}
\begin{aligned}
m_{\mathrm{e}}/m_\mu &= 0.0048\;, & m_\mu/m_\tau &= 0.0565\; , \quad& \delta_{\CP}^\ell/\pi &= 1.606\,\mathrm{rad}\;, \\
\sin^2\theta_{12} &= 0.341\;,           & \sin^2\theta_{13} &= 0.0222\;, \quad& \sin^2\theta_{23} &= 0.467\;,\\
m_1 &= 28\,\mathrm{meV}\;, & m_2 &= 30\,\mathrm{meV}\;, & m_3 &= 57\,\mathrm{meV}\;,\\
\eta_1/\pi &= 1.08\,\mathrm{rad}\;, & \eta_2 &= 0.15\,\mathrm{rad}\;.
\end{aligned}
\label{eq:BestFitFlavorObservables}
\end{equation}
Here, $m_i$ denote the neutrino masses and $\eta_i$ denote the Majorana phases. Moreover, the sum of the neutrino masses, the effective electron-neutrino mass and the effective majorana mass is given by
\begin{equation}
\sum m_\nu = 116\,\mathrm{meV} \;, \qquad
m_{\nu_{e}}^{\mathrm{eff}} = 30\,\mathrm{meV}\,, \qquad\text{and}\qquad
m_{ee} = 28\,\mathrm{meV}\,.
\end{equation}
We see that the constraints $m^{\rm eff}_{\nu_e}<0.45~{\rm eV}$  from from KATRIN ~\cite{KATRIN:2024cdt}, $m_{ee}<35~{\rm meV}$ from KamLAND-ZEN~\cite{KamLAND-Zen:2022tow}, and $\Sigma m_\nu<0.12~{\rm eV}$ from Planck~\cite{Planck:2018vyg} are all satisfied for this model.

\begin{figure}[h]
    \centering
    \includegraphics[width = \linewidth]{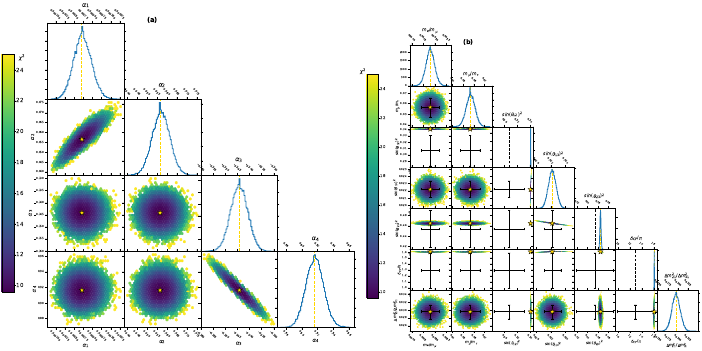}
    \caption{\textbf{Corner plots for Markov-Chain Monte-Carlo exploration around the best-fit minimum for a particular model found in the $T_{19}\times \mathbb{Z}_4$}. The best fit form the search is shown with the star indicating the best fit point. Panel \textbf{(a)} shows the $\chi^2$ distribution and correlations of the model's parameters, panel \textbf{(b)} shows the distribution of the model's predictions, along with the corresponding predictions with $2\sigma$ error bars. }
    \label{fig:CornerPlot}
\end{figure}

A Markov-Chain Monte-Carlo is performed around the minima, to explore correlations and map the $\chi^2$ distribution. In \Cref{fig:CornerPlot}\textcolor{blue}{a},  the fitted parameters and their correlations are shown. While $\alpha_3$ and $\alpha_4$ are found to be $\mathcal{O}(1)$, $\alpha_1$ and $\alpha_2$ are not. These values are attributed to the requirement that $\alpha_1$ and $\alpha_2$ generate the hierarchical masses of the charged leptons.
In \cref{fig:CornerPlot}\textbf{b}, the resulting fits to the observables are shown along with the experimentally measured values with $2\sigma$ error bars. While this model is simple and predictive, it struggles to fit $\theta_{12} $ and $\theta_{23}$ without eroding the goodness of fit of the remaining measurements. While not implemented here, it is possible that the diagnostics shown in this figure could be used as features for further explorations by \ac{AMBer}. For example, one could penalize \ac{AMBer} for fine-tuning parameters, or use correlations as feedback to find better models.

\section{Conclusions}
\label{sec:Conclusions}

Particle physics model-building has traditionally relied on a physicist's intuition to explore fields and the symmetries they obey to describe the Universe we observe. However, the model space is vast and a majority of it remains unexplored. \ac{AMBer} is presented as an AI-assisted framework to explore the high-dimensional discrete space of neutrino flavor models. It uses reinforcement learning and high-performance computing to efficiently navigate the model space using an automated end-to-end pipeline. Existing scientific software was used and redesigned, and additional software was developed and released publicly. This pipeline constructs the superpotential, extracts the mass matrices, counts the number of free parameters and fits the model to experimental data. It uncovers models that satisfy predictions which accurately match experimental observations and introduce a minimal amount of additional free parameters. These models serve as a starting point to evaluate the viability of a particular flavor symmetry to reproduce the lepton mixing parameters. Further investigations can be done on these models to incorporate the needed features of a full model. For instance, the calculation of theoretical error bars or an appropriate scalar potential could be studied, as discussed at the end of subsection \myref{Subsec:NeutrinoModelBuilding} in the \myref{Sec:Methods} section. 

\ac{AMBer} rediscovers known model patterns in the representation assignment of neutrino flavor models based on the symmetry group $A_4\times\mathbb{Z}_N$. Additionally, it finds models based on the unexplored symmetry group $T_{19}\times \mathbb{Z}_4$. This allows study of the distributions of the particle content, representation assignments, preferred Abelian symmetry group and flavon \ac{VEV} alignment. To make this possible, \ac{AMBer} is encouraged to find models with few free parameters and a good fit to data. At the same time, it is penalized for finding unrealistic models, such as those with zero electron mass. Several example models discovered by \ac{AMBer} have been explicitly discussed in the main text and in the Appendix. In particular, a neutrino flavor model based on the group $T_{19}$ with four free parameters, three flavons and $\chi^2 \approx 10$ is presented. The complete list of inequivalent filtered models (i.e., models with $\chi^2\leq10$, $n_p\leq 7$) discovered has also been released in machine-readable format and is available on the 
\texttt{Github} repository \href{https://github.com/jake-rudolph-1/models-by-AMBer}{here}. This list demonstrates \ac{AMBer}'s ability to find accurate models with few parameters. Thus, this tool can assist a physicist by filtering through a set of symmetries and field content to provide an initial set of models which can be further analyzed.

As the design of particle theory models becomes increasingly reliant on such assistants, there are interesting future research questions in how to quantitatively define the elegance of the theory, how to perform fast and accurate hypothesis tests when comparing models with different free parameters, and how to re-design scientific software to enable such searches in even more challenging model spaces. 

From a machine learning perspective,  it would be interesting to develop approaches that enable the agent to transfer knowledge acquired in one theory space to effectively explore another theory space. The ability for the agent to dynamically choose when to run physics software will enable scaling this approach to more complex software pipelines. Future work may also consider a more efficient representation of fields and symmetries to the agent, such that the agent is not given an option to choose invalid actions or generate equivalent models with seemingly different particle representation assignments. Computational aspects of the method can be improved to further leverage parallel computing capabilities at supercomputers, increasing the rate at which \ac{AMBer} can explore large theory spaces. With the growing use of \acp{LLM} for science, \ac{AMBer} could be coupled with \acp{LLM} as reinforcement learning with physics feedback, with embedded software enhancing the reliability of scientific predictions. 

From the physics perspective, it would be interesting to identify what properties make a flavor model realistic and whether this can be determined from the distribution of the particle content, representation assignment, or flavon \ac{VEV} alignment. Another future direction is to run \ac{AMBer} on a larger and more diverse set of non-Abelian flavor symmetry groups. Our approach is expected to be applicable to any finite non-Abelian flavor symmetry, provided that the corresponding mass matrices can be computed in a reasonable time. To this end, an automated method may be required to list out all of their irreducible representations. While the testability of models was not considered here, in future work, \ac{AMBer} could also be explicitly incentivized to find models that can be ruled out by upcoming experimental results. At this stage, \ac{AMBer} should be regarded as a proof of principle demonstrating that \ac{RL} can assist model builders in performing more efficient and targeted searches. The impact of our work lies in our approach to interfacing an AI agent with physics software, while the models found in this work illustrate its potential. We envision that particle physicists in the future will be able to use AI assistants such as AMBer to rapidly explore alternative ideas for model building. Further, while this work focused on neutrino flavor model-building, the approach could be applied to other areas of flavor model-building, dark matter, or even cosmology, if the interfacing with scientific software can be scaled efficiently.

\section{Data Availability}
Models found by AMBer are available in a \texttt{Github} repository at \url{https://github.com/jake-rudolph-1/models-by-AMBer}. They are stored as ".txt" files, each with a dictionary specifying all relevant information for a unique neutrino flavor model.

\section{Code Availability}
The software developed for analyzing neutrino models within the RL pipeline has been publicly released as the \texttt{FlavorBuilder} package, installable via the Python Package Index (PyPI) at
\url{https://pypi.org/project/FlavorBuilder/}. This code makes use of GAP version 4.14.0. The RL environments were developed using \texttt{gymnasium} and run using \texttt{Stable-Baselines3}. They make further use of \texttt{FlavorPy} version 0.2.0.

\section*{Acknowledgments}
AG and JR used resources from the National Energy Research Scientific Computing Center (NERSC), a U.S. Department
of Energy Office of Science User Facility located at Lawrence Berkeley National Laboratory, for this research. We specifically thank Wahid Bhimji for the support with computing resources that were essential for this work. The work of MF was supported in part by U.S. National Science Foundation Grant PHY-2210283 and the National Science Foundation Graduate Research Fellowship Award No. DGE-1839285. The work of VKP was supported by the U.S. National Science Foundation under Grant No. PHY-2210283 and by UC-MEXUS-CONACyT grant No. CN-20-38. The work of JB was supported by DOE, High Energy Physics Research and Technology Division under grant DE-SC0023966 and the auto-encoder was trained using the Greenplanet cluster (NSF Grant No. CHE- 0840513). DW and AG are supported by the DOE Office of Science. AG thanks Josef Urban for the invitation to Prague in 2016 and fruitful discussions that seeded core ideas of this work, as well participants of the AI for theorem proving conference in 2018 such as Michael Douglas and David McAllester. We would also like to thank Mu-Chun Chen, Jonathan Feng, Xueqi Li, Xiang-Gan Liu, Michael Ratz, and Mi\v{s}a Toman for fruitful discussions. We thank Wahid Bhimji, Jim Halverson, Alexander Shmakov, Jessica Howard, Michael Ratz, Tim Tait and Shimon Whiteson for providing valuable comments on the manuscript.

\section*{Author Contributions Statement}
Using the CASRAI CRediT Contributor Roles Taxonomy: Conceptualization: MF, VKP, JR, JB, AG; Software: JR, VKP; Visualization: JB, MF, JR; Resources: AG, DW; Project administration: JR, AG; Supervision: AG, DW, Validation: JR, VKP, MF, AG, DW, JB, VG; Data curation: JR, JB; Investigation: JR, MF, JB, VKP; Formal analysis: MF; Methodology: AG, JR, VKP, MF, JB; Writing – original draft: VKP, JR, MF, AG, JB; Writing – review \& editing: DW, JR, MF, VKP, AG

\section*{Competing interests}
The authors declare no competing interests.

\appendix
\section{Supplementary Material: Methods}

\subsection{Supplementary Model-Building Details}

\subsubsection{Experimental values}

\begin{table}[H]
\caption{
\textbf{Experimental central values and $1\sigma$ uncertainties for the \ac{SM} lepton sector.}
		The data for the neutrino oscillation parameters is taken from the global analysis NuFIT v5.3~\cite{Esteban:2020cvm}
		for \ac{NO} and \ac{IO}, taking the Super-Kamiokande data into account.  NuFIT assumes the \ac{SM} with three left-handed massive neutrino flavor eigenstates.
		The charged-lepton mass ratios are obtained at the GUT scale as the average between the value for $\tan\beta = 10$ and $\tan \beta = 38$~\cite{Ross:2007az}.}
	\centering
\begin{tabular}{ll ll ll}
    \toprule
    \multicolumn{2}{c}{Charged-lepton mases} & \multicolumn{2}{c}{\ac{NO} NuFit v5.3 with SK } & \multicolumn{2}{c}{\ac{IO} NuFit v5.3 with SK} \\
    \cmidrule(lr){1-2}\cmidrule(lr){3-4}\cmidrule(lr){5-6}
    observables & best-fit values & observables & best-fit values & observables & best-fit values \\
    \midrule
    $m_\mathrm{e}/m_\mu$ & $0.0048\pm0.0002$ &
    $\Delta m_{21}^2 / \Delta m_{31}^2$ & $0.0295^{+0.0012}_{-0.0010}$ &
    $\Delta m_{21}^2 / \Delta m_{32}^2$ & $0.0298^{+0.0012}_{-0.0011}$ \\[4pt]
    
    $m_\mu/m_\tau$ & $0.0565 \pm 0.0045$ &
    $\sin^2\theta_{12}$ & $0.307^{+0.012}_{-0.011}$ &
    $\sin^2\theta_{12}$ & $0.307^{+0.012}_{-0.011}$ \\[4pt]

    & &
    $\sin^2\theta_{13}$ & $0.02224^{+0.00056}_{-0.00057}$ &
    $\sin^2\theta_{13}$ & $0.02222^{+0.00069}_{-0.00057}$ \\[4pt]
    
    & &
    $\sin^2\theta_{23}$ & $0.454^{+0.019}_{-0.016}$ &
    $\sin^2\theta_{23}$ & $0.568^{+0.016}_{-0.021}$ \\[4pt]
    
    & &
    $\delta_{\CP}^{\ell}/\pi$ & $1.289^{+0.217}_{-0.139}$ &
    $\delta_{\CP}^{\ell}/\pi$ & $1.517^{+0.133}_{-0.145}$ \\[2pt]
    \bottomrule
\end{tabular}
		\label{tab:ExpDataLeptons}
\end{table}

\subsubsection{\texorpdfstring{$A_4$ properties}{A4 properties}}
\label{app:A4}

The group $A_4 = \left( \mathbb{Z}_2\times\mathbb{Z}_2 \right)\rtimes \mathbb{Z}_3$ with 12 elements is the smallest non-abelian subgroup of $\SU{3}$~\cite{Ludl:2010bj} and it is generated by the elements $S$ and $T$ such that
\begin{equation}
    S^2 = T^3 = \left( S T \right)^3 =1\;.
\end{equation}
The $12$ elements of $A_4$ are $1, S, T, ST, T S, T^2, ST^2, STS, TST, T^2S, T ST^2, T^2ST$. The group has three singlet irreducible representations $\rep{1}$, $\rep{1'}$, $\rep{1''}$ and a triplet irreducible representation $\rep{3}$. The generators for the singlets can be represented as
\begin{align}
\label{eq:Singlets}
    \rep{1}: \quad & S ~=~ 1 \quad T ~=~ 1\;, \nonumber \\
    \rep{1'}: \quad & S ~=~ 1 \quad T ~=~ e^{i 4\pi/3} \equiv \omega^2\;, \nonumber \\
    \rep{1''}: \quad & S ~=~ 1 \quad T~=~ e^{i 2\pi/3} \equiv \omega \;,
\end{align}
while the triplet can be represented by the matrices 
\begin{align}
\label{eq:A4Triplets}
    \rep{3}: \quad & S ~=~ \begin{pmatrix}
1 & 0 & 0 \\
0 & -1 & 0 \\
0 & 0 & -1
\end{pmatrix} \quad  T~=~ \begin{pmatrix}
        0 & 1 & 0 \\
        0 & 0 & 1 \\
        1 & 0 & 0 
    \end{pmatrix}\;.
\end{align}
PyDiscrete, the Python translation of the Mathematica Discrete package \cite{Holthausen:2011vd}, has been used to obtain the Clebsch-Gordan coefficients of the tensor products. The singlet representations are irreducible representations of $\mathbb{Z}_3$. Thus, they satisfy the tensor products
\begin{align}
\label{eq:SingletProducts}
    \rep{1'}\otimes \rep{1'}& ~=~ \rep{1''}\;, \nonumber  \\
        \rep{1'}\otimes \rep{1''}& ~=~ \rep{1}\;, \nonumber  \\
        \rep{1''}\otimes \rep{1''}& ~=~ \rep{1'}\;.
\end{align}
The tensor product of a non-trivial singlet with the triplet irreducible representation is given by
\begin{align}
\label{eq:SingletAndTripletA4}
    \begin{pmatrix}\phi_1 \end{pmatrix}_{\rep{1'}} \otimes  \begin{pmatrix} \psi_1 \\ \psi_2 \\ \psi_3\end{pmatrix}_{\rep{3}} ~=~  & \begin{pmatrix} \phi_1 \psi_1 \\ \omega^2 \phi_1 \psi_2 \\ \omega\phi_1 \psi_3 \end{pmatrix}_{\rep{3}}\;,\nonumber \\
    \begin{pmatrix}\phi_1 \end{pmatrix}_{\rep{1''}} \otimes \begin{pmatrix} \psi_1 \\ \psi_2 \\ \psi_3\end{pmatrix}_{\rep{3}} ~=~  & \begin{pmatrix} \phi_1 \psi_1 \\ \omega \phi_1 \psi_2 \\ \omega^2\phi_1 \psi_3 \end{pmatrix}_{\rep{3}}\;.
\end{align}
Furthermore, the tensor product of two triplets is given by
\begin{align}
    \label{eq:TripletAndTripletA4}
    \begin{pmatrix}\phi_1 \\ \phi_2 \\ \phi_3 \end{pmatrix}_{\rep{3}}\otimes\begin{pmatrix} \psi_1 \\ \psi_2 \\ \psi_3\end{pmatrix}_{\rep{3}} ~=~   & \begin{pmatrix}\varphi_3 \psi_2 \\\varphi_1 \psi_3  \\ \phi_2 \psi_1 \end{pmatrix}_{\rep{3}^{(1)}} \oplus   \begin{pmatrix} \varphi_2 \psi_3 \\
\varphi_3 \psi_1  \\
\varphi_1 \psi_2\end{pmatrix}_{\rep{3}^{(2)}} \oplus 
         \begin{pmatrix} \frac{1}{\sqrt{3}}\left(\phi_1\psi_1 + \phi_2 \psi_2 + \phi_3 \psi_3 \right) \end{pmatrix}_{\rep{1}} \oplus \nonumber \\
         & \begin{pmatrix} \frac{2\sqrt{3}}{6} \left( \,\phi_1 \psi_1 + \omega \,\phi_2 \psi_2 +\omega^2\,\phi_3 \psi_3 \right)\end{pmatrix}_{\rep{1'}} \oplus  \begin{pmatrix} \frac{2\sqrt{3}}{6} \left( \,\phi_1 \psi_1 + \omega^2 \,\phi_2 \psi_2 +\omega\,\phi_3 \psi_3 \right)\end{pmatrix}_{\rep{1''}}\;.
\end{align}

\subsubsection{\texorpdfstring{$T_{19}$ properties}{T19 properties}}
\label{app:T19}

The group $T_{19} = \mathbb{Z}_{19} \rtimes \mathbb{Z}_{3}$ with $57$ elements is a subgroup of $\SU{3}$~\cite{Ludl:2010bj} generated by $a$ and $b$ such that
\begin{equation}
    a^3  ~=~ \Id\;, \qquad b^{19} ~=~ \Id\;, \qquad \mathrm{and}\qquad ba ~=~ a b^7\;.
\end{equation}
The $57$ elements of $T_{19}$ are of the form  $b^{m}a^{n}$ for $0 \leq m \leq 18$ and $ 0 \leq n \leq 2$. The group has three singlet irreducible representations $\rep{1}$, $\rep{1'}$, $\rep{1''}$, and 6 triplet irreducible representations $\rep{3_1}$, $\rep{\bar{3}_1}$, $\rep{3_2}$, $\rep{\bar{3}_2}$, $\rep{3_3}$, $\rep{\bar{3}_3}$. The generators for the singlets can be represented in the same way as in \cref{eq:Singlets}. The triplets can be represented by
\begin{align}
\label{eq:bTriplets}
    \rep{3_1}: \quad & b ~=~ \begin{pmatrix}
\ee^{\frac{2\pi \ii}{19}} & 0 & 0 \\
0 & \ee^{-\frac{16\pi \ii}{19}} & 0 \\
0 & 0 & \ee^{\frac{14\pi \ii}{19}}
\end{pmatrix}\;, \nonumber \\
    \rep{3_2}: \quad & b ~=~ \begin{pmatrix}
\ee^{\frac{4\pi\ii}{19}} & 0 & 0 \\
0 & \ee^{\frac{6\pi\ii}{19}} & 0 \\
0 & 0 & \ee^{-\frac{10\pi\ii}{19}}
\end{pmatrix}\;, \nonumber \\
\rep{3_3}: \quad & b ~=~ \begin{pmatrix}
\ee^{-\frac{12\pi}{19}} & 0 & 0 \\
0 & \ee^{-\frac{18\pi}{19}} & 0 \\
0 & 0 & \ee^{-\frac{8\pi}{19}}
\end{pmatrix}\;, 
\end{align}
where the representation of $b$ for $\rep{\bar{3}_1}$, $\rep{\bar{3}_2}$ and $\rep{\bar{3}_3}$ can be obtained by taking the complex conjugate of \cref{eq:bTriplets}. The element $a$ has the same matrix representation in all $6$ triplet irreducible representations and it is given by 
\begin{equation}
    a~=~ \begin{pmatrix}
        0 & 1 & 0 \\
        0 & 0 & 1 \\
        1 & 0 & 0 
    \end{pmatrix}\;.
\end{equation}
Similarly to \Cref{app:A4}, PyDiscrete has been used to obtain the Clebsch-Gordan coefficients of the tensor products. The singlet representations are irreducible representations of $\mathbb{Z}_3$. Thus, the tensor products between singlet irreducible representations satisfy the same relations as in \cref{eq:SingletProducts}. The tensor product between any triplet of $T_{19}$ and a singlet satisfies \cref{eq:SingletAndTripletA4}. Furthermore, the only triplet products that have a trivial singlet in their tensor decomposition are
\begin{align}
    \begin{pmatrix}\phi_1 \\ \phi_2 \\ \phi_3 \end{pmatrix}_{\rep{3_1}}\otimes\begin{pmatrix} \psi_1 \\ \psi_2 \\ \psi_3\end{pmatrix}_{\rep{\bar{3}_1}} ~=~   & \begin{pmatrix} \phi_1 \psi_2 \\ \phi_2 \psi_3 \\ \phi_3 \psi_1\end{pmatrix}_{\rep{3_3}}\oplus   \begin{pmatrix} \phi_2 \psi_2 \\ \phi_3 \psi_3 \\ \phi_1 \psi_1\end{pmatrix}_{\rep{\bar{3}_3}} \oplus 
         \begin{pmatrix} \frac{1}{\sqrt{3}}\left( \phi_2\psi_1 + \phi_3 \psi_2 + \phi_1 \psi_3\right) \end{pmatrix}_{\rep{1}} \oplus \nonumber \\
         & \begin{pmatrix} \frac{1}{\sqrt{3}}\left(\phi_1 \psi_3 + \omega \phi_2 \psi_1 + \omega^2 \phi_3 \psi_2 \right)\end{pmatrix}_{\rep{1'}} \oplus  \begin{pmatrix} \frac{1}{\sqrt{3}}\left(\phi_1 \psi_3 + \omega^2 \phi_2 \psi_1 + \omega \phi_3 \psi_2 \right)\end{pmatrix}_{\rep{1''}}
\end{align}
\begin{align}
    \begin{pmatrix}\phi_1 \\ \phi_2 \\ \phi_3 \end{pmatrix}_{\rep{3_2}}\otimes\begin{pmatrix} \psi_1 \\ \psi_2 \\ \psi_3\end{pmatrix}_{\rep{\bar{3}_2}} ~=~   & \begin{pmatrix} \phi_2 \psi_3 \\ \phi_3 \psi_1 \\ \phi_1 \psi_2\end{pmatrix}_{\rep{3_1}}\oplus   \begin{pmatrix} \phi_2 \psi_2 \\ \phi_3 \psi_3 \\ \phi_1 \psi_1\end{pmatrix}_{\rep{\bar{3}_1}} \oplus 
         \begin{pmatrix} \frac{1}{\sqrt{3}}\left( \phi_2\psi_1 + \phi_3 \psi_2 + \phi_1 \psi_3\right) \end{pmatrix}_{\rep{1}} \oplus \nonumber \\
         & \begin{pmatrix} \frac{1}{\sqrt{3}}\left(\phi_1 \psi_3 + \omega \phi_2 \psi_1 + \omega^2 \phi_3 \psi_2 \right)\end{pmatrix}_{\rep{1'}} \oplus  \begin{pmatrix} \frac{1}{\sqrt{3}}\left(\phi_1 \psi_3 + \omega^2 \phi_2 \psi_1 + \omega \phi_3 \psi_2 \right)\end{pmatrix}_{\rep{1''}}
\end{align}
\begin{align}
    \begin{pmatrix}\phi_1 \\ \phi_2 \\ \phi_3 \end{pmatrix}_{\rep{3_3}}\otimes\begin{pmatrix} \psi_1 \\ \psi_2 \\ \psi_3\end{pmatrix}_{\rep{\bar{3}_3}} ~=~   & \begin{pmatrix} \phi_3 \psi_2 \\ \phi_1 \psi_3 \\ \phi_2 \psi_1\end{pmatrix}_{\rep{3_2}}\oplus   \begin{pmatrix} \phi_2 \psi_2 \\ \phi_3 \psi_3 \\ \phi_1 \psi_1\end{pmatrix}_{\rep{\bar{3}_2}} \oplus 
         \begin{pmatrix} \frac{1}{\sqrt{3}}\left( \phi_3\psi_1 + \phi_1 \psi_2 + \phi_2 \psi_3\right) \end{pmatrix}_{\rep{1}} \oplus \nonumber \\
         & \begin{pmatrix} \frac{1}{\sqrt{3}}\left(\phi_1 \psi_2 + \omega \phi_2 \psi_3 + \omega^2 \phi_3 \psi_1 \right)\end{pmatrix}_{\rep{1'}} \oplus  \begin{pmatrix} \frac{1}{\sqrt{3}}\left(\phi_1 \psi_2 + \omega^2 \phi_2 \psi_3 + \omega \phi_3 \psi_1 \right)\end{pmatrix}_{\rep{1''}}\;.
\end{align}
The rest of the Clebsch-Gordan coefficients for the different tensor products between triplets are given in \Cref{tab:clebsch_gordan}.

\begin{table}[t!]
    \centering
    \caption{\textbf{$\mathbf{T_{19} = Z_{19} \rtimes Z_3}$ Clebsch-Gordan decompositions.}}
    \renewcommand{\arraystretch}{1.5}
    \hspace*{-0.8cm}%
    \begin{tabular}{c|c}
        \hline
        $\begin{pmatrix}\phi_1 \\ \phi_2 \\ \phi_3 \end{pmatrix}_{\rep{3_1}}$ $\otimes$ $\begin{pmatrix} \psi_1 \\ \psi_2 \\ \psi_3\end{pmatrix}_{\rep{3_1}}~=~ $    $\begin{pmatrix} \phi_3 \psi_1 \\ \phi_1 \psi_2 \\ \phi_2 \psi_3\end{pmatrix}_{\rep{\bar{3}_1}}\oplus$   $\begin{pmatrix} \phi_1 \psi_3 \\ \phi_2 \psi_1 \\ \phi_3 \psi_2\end{pmatrix}_{\rep{\bar{3}_1}} \oplus$  $\begin{pmatrix} \phi_1 \psi_1 \\ \phi_2 \psi_2 \\ \phi_3 \psi_3\end{pmatrix}_{\rep{3_2}} $  & $\begin{pmatrix}\phi_1 \\ \phi_2 \\ \phi_3 \end{pmatrix}_{\rep{3_1}}$ $\otimes$ $\begin{pmatrix} \psi_1 \\ \psi_2 \\ \psi_3\end{pmatrix}_{\rep{3_2}}~=~ $  $\begin{pmatrix} \phi_3 \psi_3 \\ \phi_1 \psi_1 \\ \phi_2 \psi_2\end{pmatrix}_{\rep{3_2}}\oplus$ $\begin{pmatrix} \phi_2 \psi_1 \\ \phi_3 \psi_2 \\ \phi_1 \psi_3\end{pmatrix}_{\rep{3_3}}\oplus$ $\begin{pmatrix} \phi_1 \psi_2 \\ \phi_2 \psi_3 \\ \phi_3 \psi_1\end{pmatrix}_{\rep{\bar{3}_3}}$\\
        $\begin{pmatrix}\phi_1 \\ \phi_2 \\ \phi_3 \end{pmatrix}_{\rep{3_1}}$ $\otimes$ $\begin{pmatrix} \psi_1 \\ \psi_2 \\ \psi_3\end{pmatrix}_{\rep{3_3}}~=~ $  $\begin{pmatrix} \phi_3 \psi_1 \\ \phi_1 \psi_2 \\ \phi_2 \psi_3\end{pmatrix}_{\rep{3_1}}\oplus$ $\begin{pmatrix} \phi_2 \psi_2 \\ \phi_3 \psi_3 \\ \phi_1 \psi_1\end{pmatrix}_{\rep{3_2}}\oplus$ $\begin{pmatrix} \phi_1 \psi_3 \\ \phi_2 \psi_1 \\ \phi_3 \psi_2\end{pmatrix}_{\rep{\bar{3}_2}}$ & $\begin{pmatrix}\phi_1 \\ \phi_2 \\ \phi_3 \end{pmatrix}_{\rep{3_1}}$ $\otimes$ $\begin{pmatrix} \psi_1 \\ \psi_2 \\ \psi_3\end{pmatrix}_{\rep{\bar{3}_2}}~=~ $  $\begin{pmatrix} \phi_2 \psi_1 \\ \phi_3 \psi_2 \\ \phi_1 \psi_3\end{pmatrix}_{\rep{\bar{3}_1}}\oplus$ $\begin{pmatrix} \phi_2 \psi_2 \\ \phi_3 \psi_3 \\ \phi_1 \psi_1\end{pmatrix}_{\rep{\bar{3}_2}}\oplus$ $\begin{pmatrix} \phi_3 \psi_1 \\ \phi_1 \psi_2 \\ \phi_2 \psi_3\end{pmatrix}_{\rep{\bar{3}_3}}$ \\
        $\begin{pmatrix}\phi_1 \\ \phi_2 \\ \phi_3 \end{pmatrix}_{\rep{3_1}}$ $\otimes$ $\begin{pmatrix} \psi_1 \\ \psi_2 \\ \psi_3\end{pmatrix}_{\rep{\bar{3}_3}}~=~ $  $\begin{pmatrix} \phi_2 \psi_3 \\ \phi_3 \psi_1 \\ \phi_1 \psi_2\end{pmatrix}_{\rep{3_1}}\oplus$ $\begin{pmatrix} \phi_3 \psi_2 \\ \phi_1 \psi_3 \\ \phi_2 \psi_1\end{pmatrix}_{\rep{3_3}}\oplus$ $\begin{pmatrix} \phi_3 \psi_3 \\ \phi_1 \psi_1 \\ \phi_2 \psi_2\end{pmatrix}_{\rep{\bar{3}_2}}$ & $\begin{pmatrix}\phi_1 \\ \phi_2 \\ \phi_3 \end{pmatrix}_{\rep{\bar{3}_1}}$ $\otimes$ $\begin{pmatrix} \psi_1 \\ \psi_2 \\ \psi_3\end{pmatrix}_{\rep{\bar{3}_1}}~=~ $  $\begin{pmatrix} \phi_2 \psi_1 \\ \phi_3 \psi_2 \\ \phi_1 \psi_3\end{pmatrix}_{\rep{3_1}}\oplus$ $\begin{pmatrix} \phi_1 \psi_2 \\ \phi_2 \psi_3 \\ \phi_3 \psi_1\end{pmatrix}_{\rep{3_1}}\oplus$ $\begin{pmatrix} \phi_1 \psi_1 \\ \phi_2 \psi_2 \\ \phi_3 \psi_3\end{pmatrix}_{\rep{\bar{3}_2}}$ \\
        $\begin{pmatrix}\phi_1 \\ \phi_2 \\ \phi_3 \end{pmatrix}_{\rep{\bar{3}_1}}$ $\otimes$ $\begin{pmatrix} \psi_1 \\ \psi_2 \\ \psi_3\end{pmatrix}_{\rep{3_2}}~=~ $  $\begin{pmatrix} \phi_3 \psi_1 \\ \phi_1 \psi_2 \\ \phi_2 \psi_3\end{pmatrix}_{\rep{3_1}}\oplus$ $\begin{pmatrix} \phi_3 \psi_2 \\ \phi_1 \psi_3 \\ \phi_2 \psi_1\end{pmatrix}_{\rep{3_2}}\oplus$ $\begin{pmatrix} \phi_3 \psi_3 \\ \phi_1 \psi_1 \\ \phi_2 \psi_2\end{pmatrix}_{\rep{3_3}}$ & $\begin{pmatrix}\phi_1 \\ \phi_2 \\ \phi_3 \end{pmatrix}_{\rep{\bar{3}_1}}$ $\otimes$ $\begin{pmatrix} \psi_1 \\ \psi_2 \\ \psi_3\end{pmatrix}_{\rep{3_3}}~=~ $  $\begin{pmatrix} \phi_2 \psi_3 \\ \phi_3 \psi_1 \\ \phi_1 \psi_2\end{pmatrix}_{\rep{\bar{3}_1}}\oplus$ $\begin{pmatrix} \phi_1 \psi_1 \\ \phi_2 \psi_2 \\ \phi_3 \psi_3\end{pmatrix}_{\rep{3_2}}\oplus$ $\begin{pmatrix} \phi_1 \psi_3 \\ \phi_2 \psi_1 \\ \phi_3 \psi_2\end{pmatrix}_{\rep{\bar{3}_3}}$ \\
        $\begin{pmatrix}\phi_1 \\ \phi_2 \\ \phi_3 \end{pmatrix}_{\rep{\bar{3}_1}}$ $\otimes$ $\begin{pmatrix} \psi_1 \\ \psi_2 \\ \psi_3\end{pmatrix}_{\rep{\bar{3}_2}}~=~ $  $\begin{pmatrix} \phi_1 \psi_2 \\ \phi_2 \psi_3 \\ \phi_3 \psi_1\end{pmatrix}_{\rep{3_3}}\oplus$ $\begin{pmatrix} \phi_3 \psi_3 \\ \phi_1 \psi_1 \\ \phi_2 \psi_2\end{pmatrix}_{\rep{\bar{3}_2}}\oplus$ $\begin{pmatrix} \phi_3 \psi_2 \\ \phi_1 \psi_3 \\ \phi_2 \psi_1\end{pmatrix}_{\rep{\bar{3}_3}}$ &  $\begin{pmatrix}\phi_1 \\ \phi_2 \\ \phi_3 \end{pmatrix}_{\rep{\bar{3}_1}}$ $\otimes$ $\begin{pmatrix} \psi_1 \\ \psi_2 \\ \psi_3\end{pmatrix}_{\rep{\bar{3}_3}}~=~ $  $\begin{pmatrix} \phi_3 \psi_3 \\ \phi_1 \psi_1 \\ \phi_2 \psi_2 \end{pmatrix}_{\rep{\bar{3}_1}}\oplus$ $\begin{pmatrix} \phi_2 \psi_3 \\ \phi_3 \psi_1 \\ \phi_1 \psi_2\end{pmatrix}_{\rep{3_2}}\oplus$ $\begin{pmatrix} \phi_2 \psi_1 \\ \phi_3 \psi_2 \\ \phi_1 \psi_3\end{pmatrix}_{\rep{\bar{3}_2}}$ \\
        $\begin{pmatrix}\phi_1 \\ \phi_2 \\ \phi_3 \end{pmatrix}_{\rep{3_2}}$ $\otimes$ $\begin{pmatrix} \psi_1 \\ \psi_2 \\ \psi_3\end{pmatrix}_{\rep{3_2}}~=~ $  $\begin{pmatrix} \phi_3 \psi_1 \\ \phi_1 \psi_2 \\ \phi_2 \psi_3 \end{pmatrix}_{\rep{\bar{3}_2}}\oplus$ $\begin{pmatrix} \phi_1 \psi_3 \\ \phi_2 \psi_1 \\ \phi_3 \psi_2 \end{pmatrix}_{\rep{\bar{3}_2}}\oplus$ $\begin{pmatrix} \phi_1 \psi_1 \\ \phi_2 \psi_2 \\ \phi_3 \psi_3 \end{pmatrix}_{\rep{\bar{3}_3}}$ & $\begin{pmatrix}\phi_1 \\ \phi_2 \\ \phi_3 \end{pmatrix}_{\rep{3_2}}$ $\otimes$ $\begin{pmatrix} \psi_1 \\ \psi_2 \\ \psi_3\end{pmatrix}_{\rep{3_3}}~=~ $  $\begin{pmatrix} \phi_3 \psi_1 \\ \phi_1 \psi_2 \\ \phi_2 \psi_3 \end{pmatrix}_{\rep{\bar{3}_1}}\oplus$ $\begin{pmatrix} \phi_2 \psi_2 \\ \phi_3 \psi_3 \\ \phi_1 \psi_1 \end{pmatrix}_{\rep{3_3}}\oplus$ $\begin{pmatrix} \phi_2 \psi_1 \\ \phi_3 \psi_2 \\ \phi_1 \psi_3 \end{pmatrix}_{\rep{\bar{3}_2}}$ \\
        $\begin{pmatrix}\phi_1 \\ \phi_2 \\ \phi_3 \end{pmatrix}_{\rep{3_2}}$ $\otimes$ $\begin{pmatrix} \psi_1 \\ \psi_2 \\ \psi_3\end{pmatrix}_{\rep{\bar{3}_3}}~=~ $  $\begin{pmatrix} \phi_3 \psi_2 \\ \phi_1 \psi_3 \\ \phi_2 \psi_1 \end{pmatrix}_{\rep{3_1}}\oplus$ $\begin{pmatrix} \phi_1 \psi_2 \\ \phi_2 \psi_3 \\ \phi_3 \psi_1 \end{pmatrix}_{\rep{\bar{3}_1}}\oplus$ $\begin{pmatrix} \phi_3 \psi_3 \\ \phi_1 \psi_1 \\ \phi_2 \psi_2 \end{pmatrix}_{\rep{\bar{3}_3}}$ & $\begin{pmatrix}\phi_1 \\ \phi_2 \\ \phi_3 \end{pmatrix}_{\rep{3_3}}$ $\otimes$ $\begin{pmatrix} \psi_1 \\ \psi_2 \\ \psi_3\end{pmatrix}_{\rep{3_3}}~=~ $ $\begin{pmatrix} \phi_2 \psi_2 \\ \phi_3 \psi_3 \\ \phi_1 \psi_1 \end{pmatrix}_{\rep{3_1}}\oplus$ $\begin{pmatrix} \phi_2 \psi_1 \\ \phi_3 \psi_2 \\ \phi_1 \psi_3 \end{pmatrix}_{\rep{\bar{3}_3}}\oplus$ $\begin{pmatrix} \phi_1 \psi_2 \\ \phi_2 \psi_3 \\ \phi_3 \psi_1 \end{pmatrix}_{\rep{\bar{3}_3}}$ \\
        $\begin{pmatrix}\phi_1 \\ \phi_2 \\ \phi_3 \end{pmatrix}_{\rep{3_3}}$ $\otimes$ $\begin{pmatrix} \psi_1 \\ \psi_2 \\ \psi_3\end{pmatrix}_{\rep{\bar{3}_2}}~=~ $ $\begin{pmatrix} \phi_3 \psi_2 \\ \phi_1 \psi_3 \\ \phi_2 \psi_1 \end{pmatrix}_{\rep{3_1}}\oplus$ $\begin{pmatrix} \phi_2 \psi_3 \\ \phi_3 \psi_1 \\ \phi_1 \psi_2 \end{pmatrix}_{\rep{\bar{3}_1}}\oplus$ $\begin{pmatrix} \phi_3 \psi_3 \\ \phi_1 \psi_1 \\ \phi_2 \psi_2 \end{pmatrix}_{\rep{\bar{3}_3}}$ &  $\begin{pmatrix}\phi_1 \\ \phi_2 \\ \phi_3 \end{pmatrix}_{\rep{\bar{3}_2}}$ $\otimes$ $\begin{pmatrix} \psi_1 \\ \psi_2 \\ \psi_3\end{pmatrix}_{\rep{\bar{3}_2}}~=~ $ $\begin{pmatrix} \phi_2 \psi_1 \\ \phi_3 \psi_2 \\ \phi_1 \psi_3 \end{pmatrix}_{\rep{3_2}}\oplus$ $\begin{pmatrix} \phi_1 \psi_2 \\ \phi_2 \psi_3 \\ \phi_3 \psi_1 \end{pmatrix}_{\rep{3_2}}\oplus$ $\begin{pmatrix} \phi_1 \psi_1 \\ \phi_2 \psi_2 \\ \phi_3 \psi_3 \end{pmatrix}_{\rep{3_3}}$\\
        $\begin{pmatrix}\phi_1 \\ \phi_2 \\ \phi_3 \end{pmatrix}_{\rep{\bar{3}_2}}$ $\otimes$ $\begin{pmatrix} \psi_1 \\ \psi_2 \\ \psi_3\end{pmatrix}_{\rep{\bar{3}_3}}~=~ $ $\begin{pmatrix} \phi_1 \psi_1 \\ \phi_2 \psi_2 \\ \phi_3 \psi_3 \end{pmatrix}_{\rep{3_1}}\oplus$ $\begin{pmatrix} \phi_3 \psi_1 \\ \phi_1 \psi_2 \\ \phi_2 \psi_3 \end{pmatrix}_{\rep{3_2}}\oplus$ $\begin{pmatrix} \phi_3 \psi_2 \\ \phi_1 \psi_3 \\ \phi_2 \psi_1 \end{pmatrix}_{\rep{\bar{3}_3}}$ & $\begin{pmatrix}\phi_1 \\ \phi_2 \\ \phi_3 \end{pmatrix}_{\rep{\bar{3}_3}}$ $\otimes$ $\begin{pmatrix} \psi_1 \\ \psi_2 \\ \psi_3\end{pmatrix}_{\rep{\bar{3}_3}}~=~ $ $\begin{pmatrix} \phi_1 \psi_1 \\ \phi_2 \psi_2 \\ \phi_3 \psi_3 \end{pmatrix}_{\rep{\bar{3}_1}}\oplus$ $\begin{pmatrix} \phi_3 \psi_1 \\ \phi_1 \psi_2 \\ \phi_2 \psi_3 \end{pmatrix}_{\rep{3_3}}\oplus$ $\begin{pmatrix} \phi_1 \psi_3 \\ \phi_2 \psi_1 \\ \phi_3 \psi_2 \end{pmatrix}_{\rep{3_3}}$\\
        \hline
    \end{tabular}
    \label{tab:clebsch_gordan}
\end{table}

\subsection{Supplementary RL Details}

\subsubsection{Reward Function} \label{app:RewardFunctionPlots}
In order to better understand the structure of the reward function, one can plot each term over a range of possible inputs with different variable targets. For example, the top left panel of \Cref{fig:RewardShaping} demonstrates how the hyperbolic tangent structure of the $R_\chi$ term (see Eq. 7 of the “Methods” section) in the reward yields a region where the goodness of fit is well differentiated before approaching a minimum value for poor fits. This becomes stricter over the course of the run as the target changes. The same is shown for the $R_p$ term (see Eq. 8 of the “Methods” section) in the top right panel, but the logarithmic structure of this term falls off more slowly. Finally, the bottom panel shows how different models evolve over the course of the run. A relatively poor fit may be given a high reward early on to encourage the agent to find valid models. Similarly, there may be a region where a non-predictive model with a good fit is rewarded to teach the agent to prioritize good fits, until late in the run it is forced to differentiate more strongly on number of parameters. These plots help figure out how to tune the reward function to properly value a particular kind of model while discriminating against others. $\chi^2_{target}$ and $n_p^{target}$ shift every 8000 steps per environment, while $N_{target}$ shifts eveery 4000 steps to allow more changes over the course of the run. The different target values are displayed in \Cref{tab:RewardSched}, along with other hyperparameters of the reward function. Recall that the actual reward is normalized by a factor of 10. Coefficients which are the same for each run are not included in the table. They are: $c_{inv}=10$, $c_{scale}=5$, and $c_{shift}=4$.

\begin{figure}[t!]
    \centering
        \includegraphics[width=1.0\textwidth]{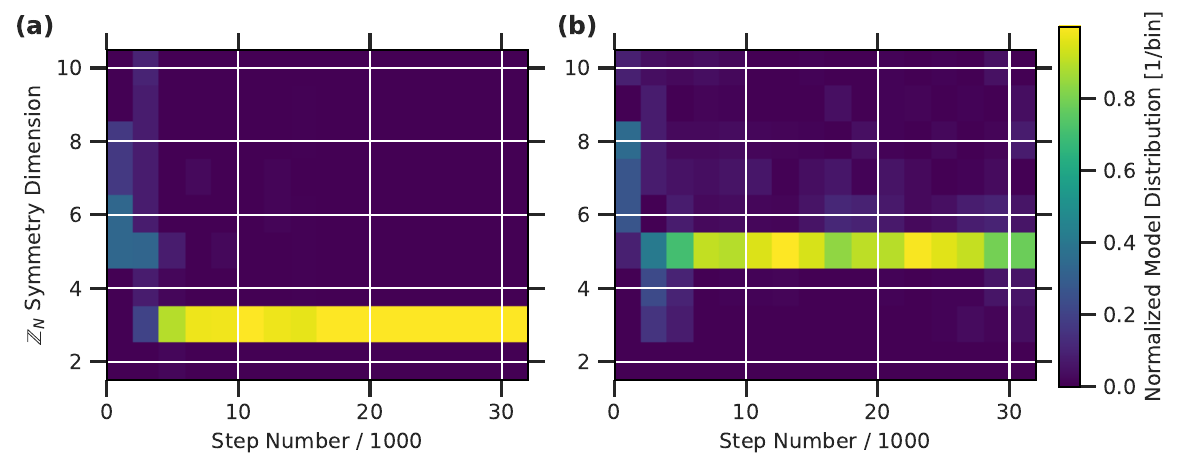}
    
    \caption{\textbf{Impact of $R_{\mathbb{Z}}$ reward term.} Distribution of the order of $\mathbb{Z}_N$ versus training step, without (panel \textbf{(a)}) and with (panel \textbf{(b)}) the $R_{\mathbb{Z}}$ reward term that encourages higher orders in models. The comparison between the two panels show how the distribution of found models change by introducing the $R_{\mathbb{Z}}$ reward term. \Cref{tab:RewardSched} describes how the $\mathbb{Z}_N$ target changes over time.}
    \label{fig:ZEvolution}
\end{figure}

\begin{figure}[t!]
    \centering
\includegraphics[width=\textwidth]{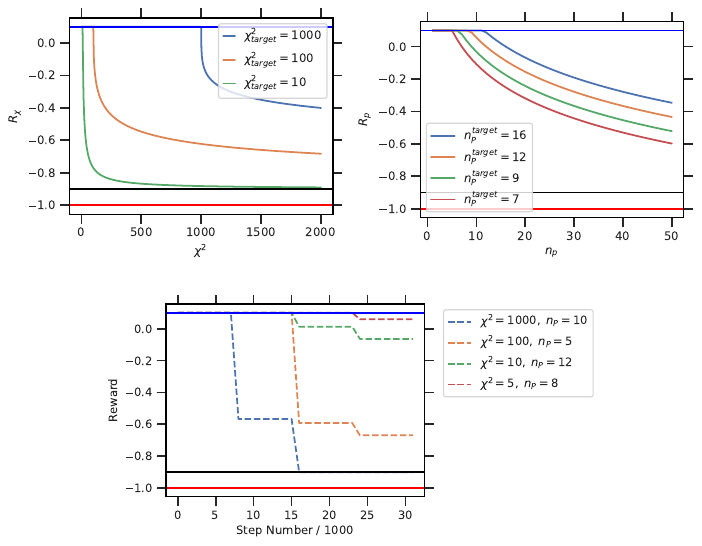}
    \caption{\textbf{Reward function evolution.} Evolution of the reward function for different targets $\chi^2_{\mathrm{target}}$ and $n_p^{\mathrm{target}}$ terms of Eq. 6 of the “Methods” section. As the targets shift, the requirements for a terminal state become more stringent, allowing the agent to focus on finding valid models early and filtered models late. In all plots, the solid blue line indicates the threshold, the solid black line the invalid model penalty $c_{\mathrm{rank}}$, and the red line the invalid action penalty $c_{\mathrm{inv}}$. Any reward over the threshold is boosted to 1, and any penalty for a valid model below the invalid model penalty is set to the invalid model penalty. Panel \textbf{(a)} shows the evolution of $R_{\chi}$ with different target values $\chi^2_{\mathrm{target}}$, panel \textbf{(b)} shows the evolution of $R_p$ with different target values $n_p^{\mathrm{target}}$, and panel \textbf{(c)} shows the total reward evolution for four models with a fixed $\chi^2$ and $n_p$.}
    \label{fig:RewardShaping}
\end{figure}

The efficacy of the $R_{\mathbb{Z}}$ term can be evaluated by tracking the distribution of the Abelian symmetry order $N$. \Cref{fig:ZEvolution} shows the evolution of this distribution among the models with $\chi^2 \leq 10$ and $ n_{p} \leq 7$ with and without $R_{\mathbb{Z}}$. As expected, the agent prefers a low order when $R_{\mathbb{Z}}$ is excluded, but more effectively explores higher orders when $R_{\mathbb{Z}}$ is included. Although the penalty term ultimately biases the agent towards $N = 5$, this term can be tuned through scheduling to more evenly search the space. 

\subsubsection{RL Training Analysis} \label{app:RLTrainingPlots}

\begin{table}[t!]
	\centering
     \caption{
                    \textbf{AMBer action space}. Description of actions that can be taken on leptons and Higgses, flavons, and globally. 
		      }
		\label{tab:ActionSpace}
	\begin{tabular}{lll}
		\toprule
		Leptons and Higgses                      & Flavons                     & Global \\
		\midrule
		  Change non-Abelian representation &  Change non-Abelian representation & Change Abelian symmetry order\\
		Change Abelian charge of a single particle    & Remove or add a particle & Evaluate model \\
	    Change Abelian charge of a set of particles   &  Change Abelian charge\\
		                              &  Change VEV configuration\\
		\bottomrule
	\end{tabular}
\end{table}

\begin{figure}[t!]
    
    \centering
    \includegraphics[width=1.0\textwidth]{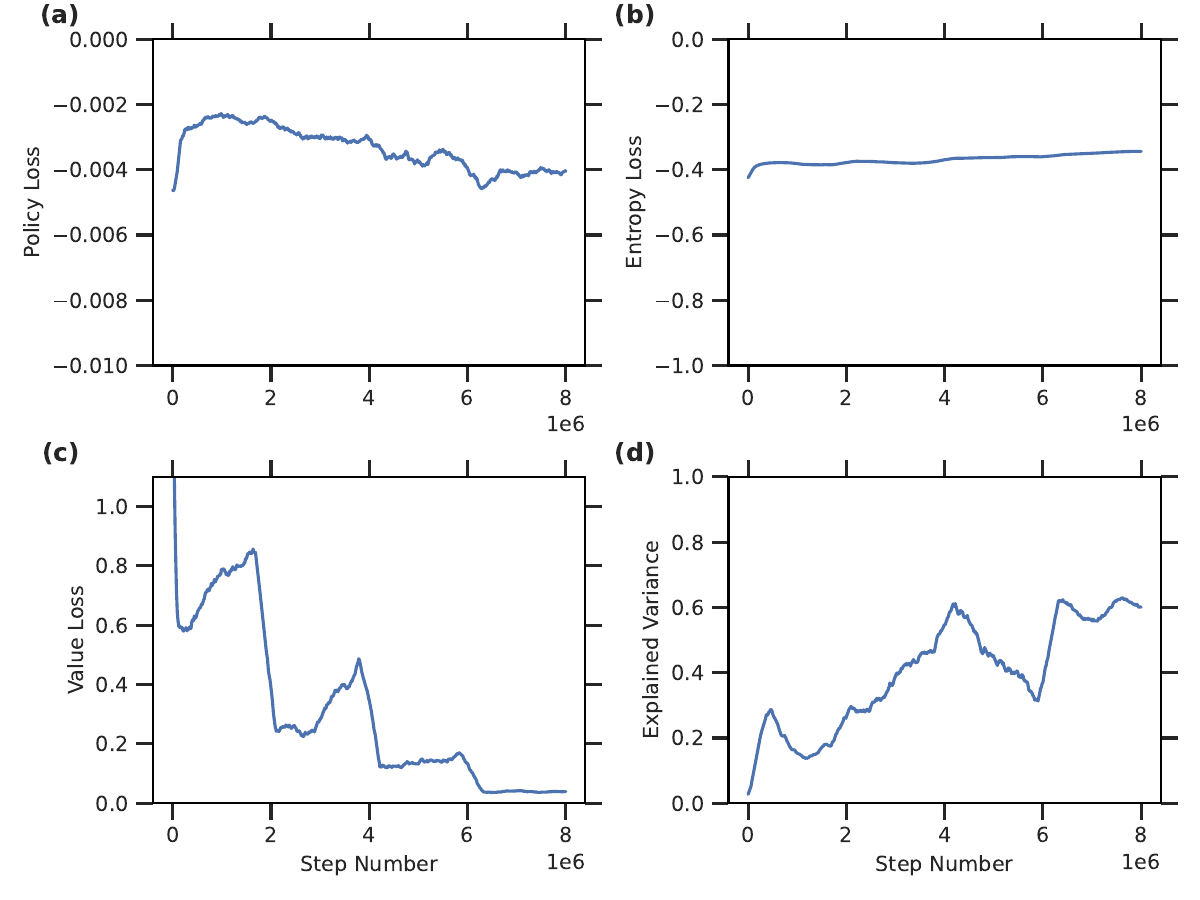}
    
    \caption{\textbf{PPO metrics.} These demonstrate how the agent is learning from an individual $A_4 \times \mathbb{Z}_4$ run. The policy loss is shown in panel \textbf{(a)}. When this loss decreases, it indicates the agent is selecting more advantageous actions. Panel \textbf{(b)} shows the entropy loss. The use of this term in the loss function prevents a more significant increase in the actual entropy. The value loss is shown in panel \textbf{(c)}. It can be seen that the value loss increases when the reward is changed by shifting targets. The explained variance is shown in panel \textbf{(d)}. The explained variance in general increases, but does not converge to one. Given the complexity of the reward function and the competing goals encoded within, it follows that the agent has a difficult time finding an optimal policy, but improves significantly over random exploration.}
    \label{fig:RLMetrics}
\end{figure}

In addition to the performance of the variables of interest in the reward function, one can look at the performance of the neural networks themselves. The three loss terms used by the neural networks are shown in \Cref{fig:RLMetrics}. In general, the loss is dominated by the value loss (panel \textbf{c}). This term oscillates over the course of the run as the performance goals change, but each time the value loss decreases, indicating the value network has learned how to predict the new reward. The explained variance (panel \textbf{d}) describes how well the variance in the predictions of the value network matches the variance in the actual return. When the explained variance is 1, the value network predictions have the same variance as the actual return, while when it is 0, the value network fails to describe any of the variance in the return. As seen in \Cref{fig:RLMetrics} panel \textbf{d}, at the start of the run, the value network has no explanatory power, while over time this rises, indicating that the value network is learning to accurately predict the reward associated with a particular state. Again, periodic rises and falls indicate how the agent adjusts to updated performance goals. Using all of these plots, one can evaluate how well the agent is learning, both in terms of the performance of the networks that compose the agent and in terms of the physics goals.

\begin{table}[t!]
\centering
\caption{\textbf{Reward function hyperparameters.} The two $A_4$ runs utilize the same $\chi^2_{\mathrm{target}}$ schedules, while the $T_{19}$ run utilizes a more lenient schedule, making it easier for the agent to focus on valid models in the larger space. $\chi^2_{target}$ and $n_p^{target}$ change every 8000 steps, while $N_{target}$ changes every 4000 steps. Hyperparameters were manually tuned to find the best performing values.}
\begin{tabular}{c c c c c c c c} 
  \toprule
    Space                     & $c_1$ & $c_2$ & $c_3$  & $c_{rank}$ & $\chi^2_{\mathrm{target}}$  & $n_p^{\mathrm{target}}$  & $N_{target}$ \\ \midrule
    $A_4 \times \mathbb{Z}_4$ &   1   &   7   &   0        &    9     & $[10^3, 10^2, 10, 10]$     & $[16, 12, 9, 7]$            & N/A  \\
    $A_4 \times \mathbb{Z}_N$ &   1   &  7.5  & 0.35     &    8     & $[10^3, 10^2, 10, 10]$     & $[18, 14, 10, 7]$           & $[4, 5, 6, 7, 8, 8, 9, 9]$  \\ 
    $T_{19} \times \mathbb{Z}_4$ & 1.5  &  5    &   0     &    8     & $[10^5, 10^4, 10^3, 10^2]$ & $[20, 16, 12, 8]$           & N/A \\
    \bottomrule
\end{tabular}
\label{tab:RewardSched}
\end{table}

\subsection{Supplementary Visualizations}

\subsubsection{Autoencoder Training and additional latent plots} \label{app:Autoencoder}
Auto-encoders map data to an intermediate latent space and back to the original space. They are trained in an unsupervised manner to learn a mapping which minimizes the distance between the original data and the mapped-and-unmapped data~\cite{Bank2023,ACKLEY1985147}.  There is no strict guarantee of interpretability or smoothness of the latent space, but when the latent space is lower-dimensional than the original data, it can be understood as a compressed version, maintaining only the critical structure and discarding degenerate information. In addition to the particle representation assignments and VEVs, calculated quantities such as the number of free parameters, information about the rank of the mass matrices, and the number of flavons for each theory are included as input features to the auto-encoder to encourage further structure. In \Cref{fig:latplot2}, a side-by-side comparison of the latent space and a histogram of the full run confirms that the agent is sufficiently thorough in its search, as we see no major regions are excluded. The hyperparameters are shown in \Cref{tab:AEHyperParameters}.

\subsubsection{Visualization in two-dimensions}
\label{subsec:Visualization}

\begin{figure}[h!]
    \centering
    \includegraphics[width=0.9 \textwidth]{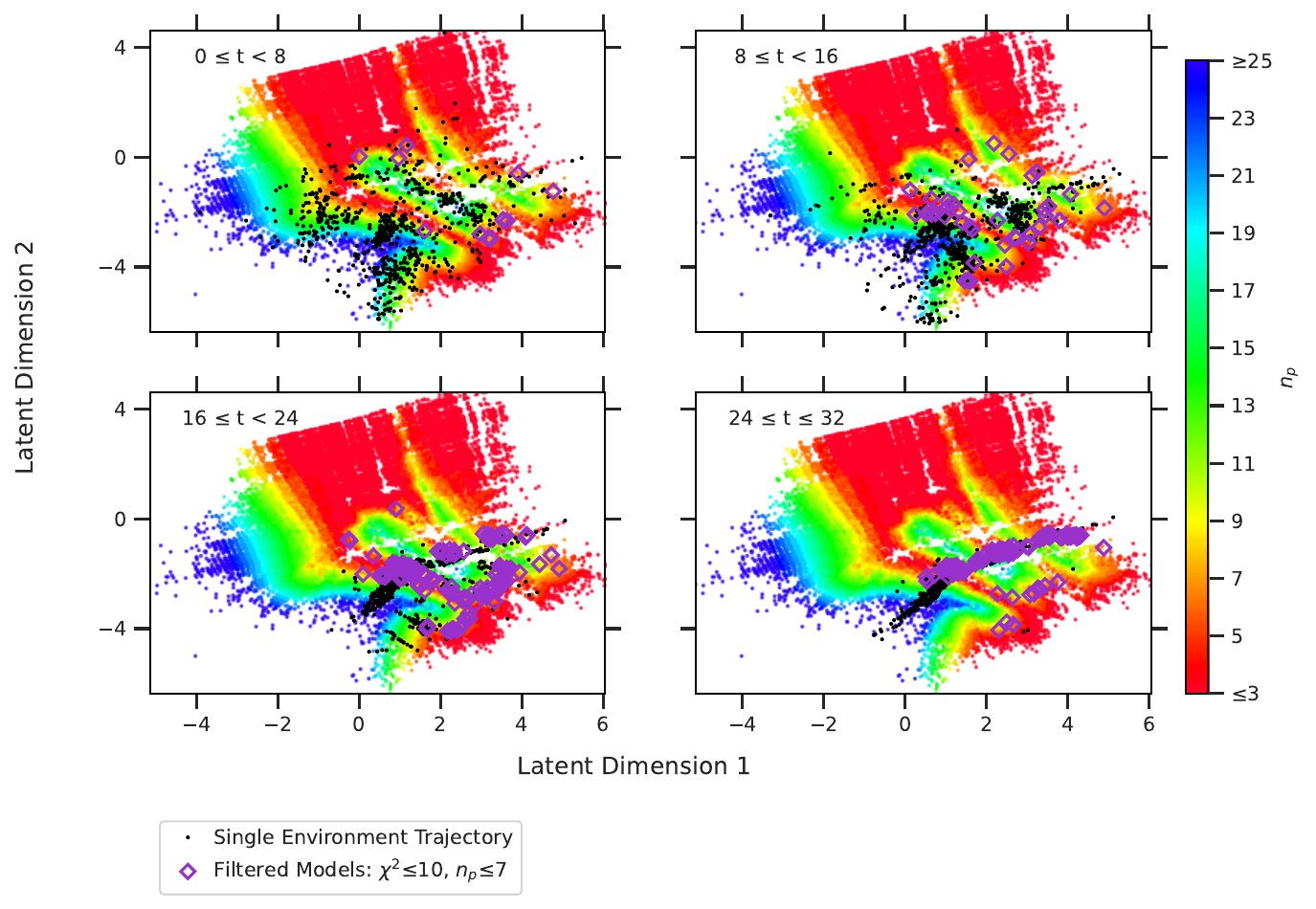}
    \caption{\textbf{Environment trajectory in the autoencoder latent space.} A single environment trajectory in the $A_4\times \mathbb{Z}_4$ latent space over the course of the full run is plotted, where the black dots represent a model that AMBer studied and the colored dots map the number of parameters across the latent space. Each panel displays the distribution of models in a different quartile of the run (where t=1 represents 1000 timesteps). Early on the agent searches more broadly throughout the space before honing in on specific promising regions. Filtered models ($\chi^2 \leq 10$ and $n_p \leq$  7) are indicated by magenta diamonds, showing that the number of promising models found increases as the agent learns to exploit the search space.
    }
    \label{fig:latplot3}
\end{figure}

While the end result of \ac{AMBer}'s search are potentially useful models, there may also be gains in understanding the path it takes as it searches, which could yield insight into the structure of the space. The high dimensionality of the theory space prevents direct visualization of \ac{AMBer}'s search trajectories. An unsupervised auto-encoder is trained to map the theory space to an abstract two-dimensional latent space, where visualization and analysis can be significantly easier ~\cite{Mutter:2019,Baretz2023-xm}. \ac{AMBer}'s search path for a single environment trajectory is also shown in \Cref{fig:latplot3}. Each panel represents 1/4 of the total search time, demonstrating how the agent shifts from exploration to exploitation as it learns to focus on regions where the most promising models cluster. Additional details on the auto-encoder training and hyperparameters can be found in \Cref{app:Autoencoder}.

\begin{figure}[t!]
    \centering
    \includegraphics[width=1.0
    \textwidth]{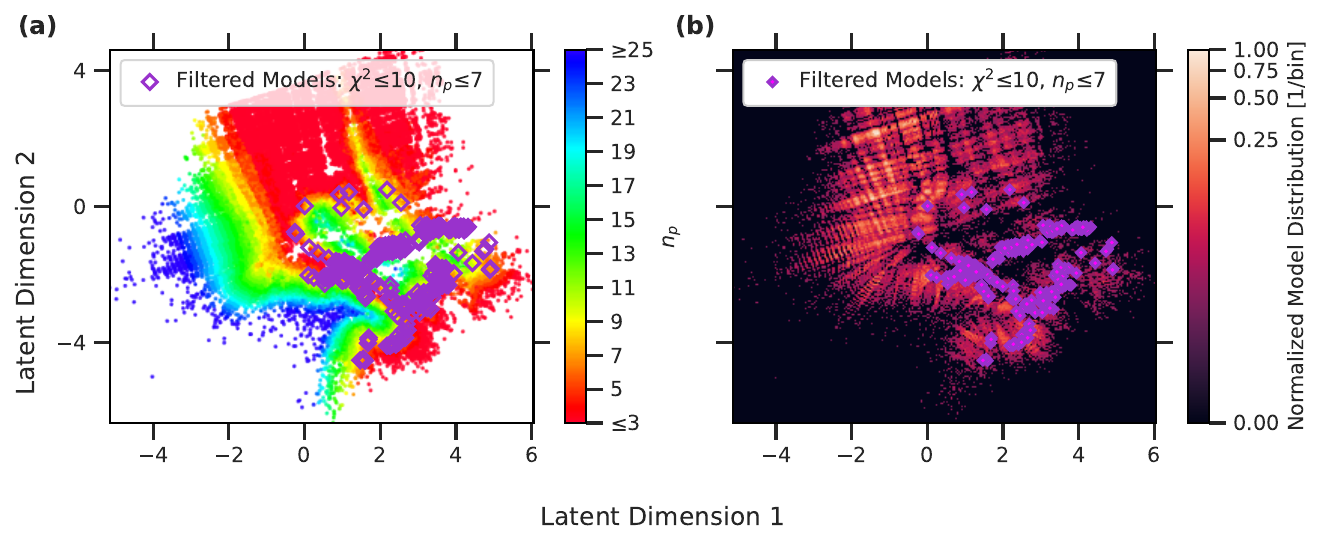}
    
    \caption{\textbf{Latent space visualization.} Panel \textbf{(a)} shows 500k background points are used to visualize the $A_4\times \mathbb{Z}_4$ latent space. Filtered models ($\chi^2 \leq 10$ and $n_p \leq$  7) are included by magenta diamonds for reference. Panel \textbf{(b)} shows a 2D histogram of the full run in the same latent space. While the agent spends the majority of its time honing in on specific regions, we can see that, in general, no significant portion of the search space is overlooked or excluded.
    }
    \label{fig:latplot2}
\end{figure}

\begin{table}[t!]
\centering
\caption{\textbf{Autoenconder hyperparameters}. For visualization, a standard (non-generative) auto-encoder is utilized, trained on $\sim$10 million models. 
}
\begin{tabular}{c c } 
  \toprule
  Parameter & $A_4\times \mathbb{Z}_4$  \\ [0.5ex] 
  \midrule
 Batch Size & 5,000 \\ 
 Learning Rate & 0.0001 \\
 Optimizer & ADAM \\  
 Num Layers & 9 \\  
 Activation Function & Leaky RELU \\
 Num Epochs & 200  \\
 \bottomrule
\end{tabular}
\label{tab:AEHyperParameters}
\end{table}

\FloatBarrier
\section{Supplementary Material: Results}


\subsection{Additional Individual Models}\label{app:AdditionalModels}
In this appendix, one model per theory space is presented to illustrate the types of models that can be found by \ac{AMBer}. First, a model that was found during the $A_4\times \mathbb{Z}_4$ is explicitly presented. The model is given by Tab. 1 in the “Methods” section. It has  5 parameters and $\chi^2=5.6$. Here $\langle\phi_1\rangle / \left( 0.1 \Lambda \right)= (0,1,0)$ and $\langle\phi_2\rangle / \left( 0.1 \Lambda \right)= (1,\omega,\omega^2)$. The mass matrices for this model are

\begin{align}
\label{eq:ExampleModela4z4_model}
    m_C &= v_d~\hat{\alpha}_C 
\begin{bmatrix}
\alpha_8 & 0 & 1 \\
0 & \alpha_8 & 1 \\
-\alpha_9 & 0 & \alpha_{8}
\end{bmatrix}, \nonumber \\
    m_M &= \Lambda \hat{\alpha}_M
\begin{bmatrix}
2 & \alpha_4\!\left(-\frac{\sqrt{3}}{6} + \frac{1}{2}\,i\right) & -\frac{\sqrt{3}}{3}\,\alpha_4\!\left(\frac{1}{2} + i\,\frac{\sqrt{3}}{2}\right) \\[6pt]
\alpha_4\!\left(-\frac{\sqrt{3}}{6} + \frac{1}{2}\,i\right) & -1 - i\,\sqrt{3} & \frac{\sqrt{3}}{3}\,\alpha_4 \\[6pt]
-\frac{\sqrt{3}}{3}\,\alpha_4\!\left(\frac{1}{2} + i\,\frac{\sqrt{3}}{2}\right) & \frac{\sqrt{3}}{3}\,\alpha_4 & -1 + i\,\sqrt{3}
\end{bmatrix}, \nonumber \\
    m_D &= v_u \hat{\alpha_D}
\begin{bmatrix}
\frac{1}{\sqrt{3}} & 0 & \frac{\alpha_6}{10}\!\left(-\frac{\sqrt{3}}{6} + \frac{1}{2}\,i\right) \\[6pt]
0 & -\frac{1}{2\sqrt{3}} + \frac{i}{2} & 0 \\[6pt]
\frac{\alpha_7}{10}\!\left(-\frac{\sqrt{3}}{6} + \frac{1}{2}\,i\right) & 0 & -\frac{1}{2\sqrt{3}} - \frac{i}{2}
\end{bmatrix}\;,
\end{align}
with $\alpha_{4}=0.0047,\alpha_6=0.71,\alpha_7=-0.86,\alpha_8=2.7\times 10^{-5},\alpha_9=0.057$.

Second, a model that was found during the $A_4\times \mathbb{Z}_N$ is explicitly presented. The model is given by \Cref{tab:a4zn_found_model}. It has 5 parameters and $\chi^2=0.87$.

\begin{table}[t!]
\centering
\caption{\textbf{A model found by AMBer in the $A_{4}\times \mathbb{Z}_N$ search}. In this case $N=5$, with 5 parameters and $\chi^2=0.87$. Here $\langle \phi_1 \rangle /(0.1\Lambda) = (1,1,1),\langle \phi_2 \rangle /(0.1\Lambda) = (1,\omega^2,\omega)$.}
\begin{tabular}{lllllllllll}
\toprule
  & $L$  & $E_1$ & $E_2$ & $E_3$& $N$ & $H_u$ & $H_d$ & $\phi_1$  & $\phi_2$ & $\phi_3$ \\ 
 \midrule
 $A_4$  & $\rep{3}$ & $\rep{1}$ & $\rep{1'}$ & $\rep{1''}$ & $\rep{3}$  & $\rep{1'}$  & $\rep{1'}$ & $\rep{3}$ & $\rep{3}$ & $\rep{1}$ \\ 
 $\mathbb{Z}_5$ & $4$ & $1$ & $1$ & $1$ & $1$ & $4$ & $1$ & $4$ & $1$ & $0$ \\
 \bottomrule
\end{tabular}
\label{tab:a4zn_found_model}
\end{table}

\begin{align}
\label{eq:ExampleModela4zn_model}
    m_C &= v_d~\hat{\alpha}_C 
\begin{bmatrix}
\alpha_8 & \alpha_9 & 1 \\
-\alpha_8 \left( \frac{1}{2} + \frac{\sqrt{3}}{2}i \right) & \sqrt{3}\alpha_9(\frac{\sqrt{3}}{6} + \frac{i}{2}) & 1 \\
\sqrt{3}\alpha_8(-\frac{\sqrt{3}}{6} + \frac{i}{2}) & \alpha_9(\frac{1}{2}-\frac{i\sqrt{3}}{2})^2 & 1
\end{bmatrix}, \nonumber \\
    m_M &= \Lambda \hat{\alpha}_M
\begin{bmatrix}
2\sqrt{3} & \alpha_4 & \alpha_4 \\
\alpha_4 & 2 \sqrt{3} & \alpha_4\\
\alpha_4 & \alpha_4 &2\sqrt{3}
\end{bmatrix}, \nonumber \\
    m_D &= v_u \hat{\alpha_D}
\begin{bmatrix}
\alpha_7 & \sqrt{3}\alpha_6(\frac{\sqrt{3}}{6}-\frac{i}{2}) & -\frac{1}{2}+\frac{i\sqrt{3}}{2} \\
-\frac{1}{2}-\frac{i\sqrt{3}}{2} & -\sqrt{3}(\frac{\sqrt{3}}{6}+\frac{i}{2}) & \alpha_6 \\
\sqrt{3}\alpha_6(\frac{\sqrt{3}}{6}+\frac{i}{2}) & 1 & \sqrt{3}\alpha_7(\frac{\sqrt{3}}{6}+\frac{i}{2})
\end{bmatrix}
\end{align}
with $\alpha_4 = -0.56, \alpha_6= 0.96, \alpha_7= -0.24, \alpha_8=0.057,$ and $ \alpha_9= 0.00027$.

Finally, we show an additional model that was found in the $T_{19}\times \mathbb{Z}_4$ search. The model is given in \Cref{tab:secondfoundt19model}, and has 5 parameters with $\chi^2=0.79$.

\begin{table}[t!]
\centering
\caption{\textbf{A model found by AMBer in the $T_{19}\times \mathbb{Z}_4$ search}. The model has  5 parameters and $\chi^2=0.79$. Here $\langle\phi_1\rangle/(0.1\Lambda) = (1,0,0)$,  $\langle\phi_2\rangle/(0.1\Lambda) = (0,1,-1) ,\langle\phi_3\rangle/(0.1\Lambda) = (1,1,1),\langle\phi_5\rangle/(0.1\Lambda) = (0,1,-1),\langle\phi_6\rangle/(0.1\Lambda) = (1,\omega^2,\omega)$.}
\begin{tabular}{lllllllllllll}
\toprule
  & $L$  & $E$ & $N$ & $H_u$ & $H_d$ & $\phi_1$ & $\phi_2$ & $\phi_3$  & $\phi_4$ & $\phi_5$ & $\phi_6$ \\ 
 \midrule
 $T_{19}$  & $\rep{\bar{3}}_2$ & $\rep{3}_2$ & $\rep{3}_2$  & $\rep{1'}$  & $\rep{1}$ & $\rep{\bar{3}}_1$ & $\rep{\bar{3}}_1$  & $\rep{\bar{3}}_2$  & $\rep{1''}$ & $\rep{3}_1$ & $\rep{3}_1$  \\ 
 $\mathbb{Z}_4$ & $3$ & $1$ & $1$ & $1$ & $2$ & $2$ & $0$ & $1$ & $2$ & $2$ & $3$  \\ \bottomrule
\end{tabular}
\label{tab:secondfoundt19model}
\end{table}

The mass matrices for this model are
\begin{align}
\label{eq:ExampleModelt19z4_model}
    m_C &= v_d~\hat{\alpha}_C 
\begin{bmatrix}
0 & \sqrt{3}\,\alpha_7\!\left(-\frac{\sqrt{3}}{6} + \frac{1}{2}\,i\right) & 1 \\[4pt]
0 & \alpha_8 & -\sqrt{3}\,\alpha_7\!\left(\frac{1}{2\sqrt{3}} + \frac{i}{2}\right) \\[4pt]
\alpha_7 & 0 & -\alpha_8
\end{bmatrix}, \nonumber \\
    m_M &= \Lambda \hat{\alpha}_M
\begin{bmatrix}
2\,\alpha_3 & \alpha_2 + \alpha_4\!\left(\frac{1}{2} - i\,\frac{\sqrt{3}}{2}\right)^2 & \alpha_2 + \alpha_4\!\left(\frac{1}{2} + i\,\frac{\sqrt{3}}{2}\right)^2 \\[4pt]
\alpha_2 + \alpha_4\!\left(\frac{1}{2} - i\,\frac{\sqrt{3}}{2}\right)^2 & 2\,\alpha_3 & \alpha_2 + \alpha_4 \\[2pt]
\alpha_2 + \alpha_4\!\left(\frac{1}{2} + i\,\frac{\sqrt{3}}{2}\right)^2 & \alpha_2 + \alpha_4 & 2\,\alpha_3 - 2
\end{bmatrix}, \nonumber \\
    m_D &= v_u \hat{\alpha_D}
\begin{bmatrix}
-\frac{1}{2} + i\,\frac{\sqrt{3}}{2} & 0 & 0 \\[2pt]
0 & -\frac{1}{2} - i\,\frac{\sqrt{3}}{2} & 0 \\[2pt]
0 & 0 & 1
\end{bmatrix}
\end{align}
with $\alpha_2= 0.44, \alpha_3= 0.38, \alpha_4= 0.13, \alpha_7 =  -0.00027,$ and $ \alpha_8 = 0.057$.

\bibliography{main}

\begin{acronym}
  \acro{CMB}{Cosmic Microwave Background}
  \acro{SM}{Standard Model}  
  \acro{PMNS}{Pontecorvo–Maki–Nakagawa–Sakata}
  \acro{VEV}{vaccum expectation value}
  \acro{TBM}{tri-bimaximal}
  \acro{MSSM}{Minimal Supersymmetric Standard Model}
  \acro{EWSB}{Electroweak Symmetry Breaking}
  \acro{RL}{reinforcement learning}
  \acro{PPO}{Proximal Policy Optimization}
  \acro{LLM}{Large Language Model}
  \acro{RGE}{Renormalization Group Equations}
  \acro{AMBer}{Autonomous Model Builder}
  \acro{NO}{Normal Ordering}
  \acro{IO}{Inverted Ordering}
\end{acronym}

\end{document}